\documentclass[journal]{IEEEtran}
\usepackage{amsmath,amsfonts}
\usepackage{algorithmic}
\usepackage{algorithm}
\usepackage{array}
\usepackage[caption=false,font=normalsize,labelfont=sf,textfont=sf]{subfig}
\usepackage{textcomp}
\usepackage{stfloats}
\usepackage{url}
\usepackage{verbatim}
\usepackage{graphicx}
\usepackage{cite}
\usepackage{mathtools}
\usepackage{threeparttable}
\usepackage{multirow}
\usepackage[colorlinks, linkcolor=red, anchorcolor=black, citecolor=green]{hyperref}

\begin{document}

\title{Task-Oriented Semantic Communication for Stereo-Vision 3D Object Detection}

\author{
	Zijian~Cao,~\IEEEmembership{Graduate Student Member,~IEEE,} 
	Hua~Zhang,~\IEEEmembership{Member,~IEEE,}
	Le~Liang,~\IEEEmembership{Member,~IEEE,}
	Haotian~Wang,
	Shi~Jin,~\IEEEmembership{Fellow,~IEEE}
	and~Geoffrey~Ye~Li,~\IEEEmembership{Fellow,~IEEE}
	
	\thanks{
		% Zijian Cao, Hua Zhang, Haotian Wang and Shi Jin are with the National Mobile Communications Research Laboratory, Southeast University, Nanjing 210096, China (e-mail: caozijian@seu.edu.cn; huazhang@seu.edu.cn; haotian\_wang@seu.edu.cn; jinshi@seu.edu.cn). 
		
		% Le Liang is with the National Mobile Communications Research Laboratory, Southeast University, Nanjing 210096, China, and also with Purple Mountain Laboratories, Nanjing 211111, China (e-mail: lliang@seu.edu.cn).
		
		% G. Y. Li is with the Department of Electrical and Electronic Engineering, Imperial College London, London SW7 2AZ, UK (e-mail: geoffrey.li@imperial.ac.uk).

		Zijian Cao, Hua Zhang, Haotian Wang and Shi Jin are with the National Mobile Communications Research Laboratory, Southeast University, Nanjing 210096, China (e-mail: caozijian@seu.edu.cn; huazhang@seu.edu.cn; haotian\_wang@seu.edu.cn; jinshi@seu.edu.cn). Zijian Cao is also with the Department of Electrical and Electronic Engineering, Imperial College London, London SW7 2AZ, UK. 
		
		Le Liang is with the National Mobile Communications Research Laboratory, Southeast University, Nanjing 210096, China, and also with Purple Mountain Laboratories, Nanjing 211111, China (e-mail: lliang@seu.edu.cn).
		
		G. Y. Li is with the Department of Electrical and Electronic Engineering, Imperial College London, London SW7 2AZ, UK (e-mail: geoffrey.li@imperial.ac.uk).
	}
}

\maketitle

\begin{abstract}
	With the development of computer vision, 3D object detection has become increasingly important in many real-world applications. Limited by the computing power of sensor-side hardware, the detection task is sometimes deployed on remote computing devices or the cloud to execute complex algorithms, which brings massive data transmission overhead. In response, this paper proposes an optical flow-driven semantic communication framework for the stereo-vision 3D object detection task. The proposed framework fully exploits the dependence of stereo-vision 3D detection on semantic information in images and prioritizes the transmission of this semantic information to reduce total transmission data sizes while ensuring the detection accuracy. Specifically, we develop an optical flow-driven module to jointly extract and recover semantics from the left and right images to reduce the loss of the left-right photometric alignment semantic information and improve the accuracy of depth inference. Then, we design a 2D semantic extraction module to identify and extract semantic meaning around the objects to enhance the transmission of semantic information in the key areas. Finally, a fusion network is used to fuse the recovered semantics, and reconstruct the stereo-vision images for 3D detection. Simulation results show that the proposed method improves the detection accuracy by nearly 70\% and outperforms the traditional method, especially for the low signal-to-noise ratio regime.
\end{abstract}


\section{Introduction}	
	\IEEEPARstart{A}{s} a promising research direction of computer vision, 3D object detection has played an important role in intelligent video surveillance, robot navigation, autonomous driving \cite{E1, E2}, etc. Based on the cameras or the laser radars (LiDARs), it can identify and locate objects in a 3D environment and provide the detection results for subsequent decision-making by intelligent agents. However, limited by the computing power on the sensor side, complex 3D detection algorithms are sometimes deployed on remote computing devices or clouds, which brings challenges in data exchange between sensors and remote devices \cite{E6, E7, E8}. Whether it is the point cloud data required for LiDAR detection methods \cite{E3} or the RGB images needed for monocular \cite{E4} or stereo-vision detection methods \cite{E5}, transmitting massive sensor data under limited communication bandwidth has always been a challenge and thus has become an important area of communication research. Meanwhile, the distortion of these sensor data will directly affect the 3D detection accuracy. 

	To solve this problem, several research focuses on optimizing communication resource allocation between sensors and remote computing devices \cite{E8, E11, E12}. By optimizing transmission power, bandwidth allocation, beamforming vectors, etc., these studies can improve the transmission efficiency of the entire system. Meanwhile, another line of research focuses on compressing the transmission data, hence reducing the communication payload sizes. For example, the variational image compression algorithm in \cite{N1} compresses the extracted features from the LiDAR point cloud. Then, the 1D convolutional autoencoder in \cite{N2, N3, N4} is trained to improve the detection performance under limited bandwidth. \textcolor{black}{However, these methods treat all parts of the transmitted data equally during compression, overlooking that different parts contribute differently to detection performance. Given this disparity in contribution, segments with a more significant impact on detection performance should be subjected to less compression to preserve critical information. In contrast, segments with less impact can be compressed at a higher ratio for more significant data reduction.} Besides, various works, e.g., teacher-student distillation method \cite{N2}, handshake communication mechanism \cite{N5}, ego evaluation cropping method \cite{N6}, etc., are proposed from difference perspectives for performance-bandwidth trade-off. These works are mainly oriented to collaborative perception tasks. The methods in \cite{N2} and \cite{N5} reduce the overall data transmission by judiciously deciding on which agent to communicate at each transmission opportunity. However, they are unable to cope with the challenge in each scheduled communication link that is tasked with delivering a large payload. The work in \cite{N6} reduces the transmitted data by cropping the transmitted sensor data based on the evaluation range of the receiver, i.e., the ego agent. However, this method requires both agents to have detection capabilities and overlapping detection areas, which is unsuitable for single sensor systems since the remote computing device usually lacks detection capabilities.
	
	To address the above challenges, semantic communication is introduced to improve the downstream task performance under limited communication bandwidth. Unlike traditional methods, semantic communication considers the meaning or semantics contained in the transmitted data, i.e., semantic information, during the transmission process. It attempts to deliver task-related semantic information over noisy channels rather than achieving bit- or symbol-level error-free transmission \cite{E15}. By designing specialized semantic and channel codecs, semantic communication systems can selectively extract and compress task-related semantic information and effectively overcome channel impairments to improve communication efficiency under limited bandwidth. Currently, semantic communication has achieved outstanding performance in single modal transmission tasks including, e.g., text \cite{N7, N8}, images \cite{N9, N10}, and speech \cite{N11, N12}, and multimodal transmission tasks \cite{N13, N14}. For the 3D object detection task, the spatial confidence-aware communication strategy in \cite{N15} compresses and transmits task-related semantic information in LiDAR point clouds or monocular images, reducing the requirement for communication bandwidth. The novel semantic communication architecture in \cite{N16} is designed for LiDAR-based 3D object detection tasks in autonomous driving. It can only extract and compress semantic information related to 3D detection from the LiDAR point cloud based on a specialized channel codec network. 
	
	However, the above works mainly focus on transmitting LiDAR point cloud data and are not readily applicable to stereo vision 3D detection tasks. Unlike LiDAR-based methods that rely on point cloud data, stereo-vision methods perform 3D position estimation based on left and right RGB images \cite{E5, E14}. Compared to the LiDAR-based methods, the stereo-vision methods offer a wider detection range, richer semantic information, and lower cost \cite{E14}. In contrast to monocular-based methods, stereo cameras provide more precise depth information through left-right photometric alignment. Therefore, stereo vision-based detection has a desirable trade-off in costs and performance and holds significant potential in addressing 3D detection tasks \cite{E14}. Motivated by this, we study the transmission of stereo-vision images for 3D object detection tasks and propose a semantic communication framework based on deep neural networks (DNNs) to improve the detection accuracy under limited communication bandwidth.
	
	The proposed semantic communication architecture attempts to address two essential requirements of data transmission in stereo-vision 3D object detection. First, the design of the transmitter needs to be lightweight enough to be deployed on the sensor-side equipment. Second, the transceiver needs to reduce the transmitted data while retaining the image semantic information to improve 3D detection under limited communication bandwidth. \textcolor{black}{Specifically, to meet the first requirement, we only perform semantic information extraction and compression at the transmitter to alleviate the computing pressure on the sensor side. Afterwards, complex feature extraction, optical flow-based image reconstruction, and 3D detection are performed later on the receiver side, which has sufficient computing resources.} For the second requirement, we study the operations of stereo-vision 3D detection and split it into two sub-processes. Resembling human eyes, the detection scheme first performs 2D object detection on a single image and then uses the photometric alignment \cite{E14} between images to infer depth information. Thus, in the first sub-process, a 2D detection network is introduced to extract the semantic information related to detected targets on a single image. In the second sub-process, a convolutional neural network (CNN)-based extraction network is proposed to jointly extract global information from the stereo images, which contains photometric alignment information and is used for depth inference. The extracted semantics are then compressed by CNNs to reduce the transmitted data. Correspondingly, two sets of recovery networks are deployed at the receiver to recover the semantic information. \textcolor{black}{The first recovery network consists of residual networks, which recover the target-related semantics. The second network is driven by an optical flow module, where optical flow \cite{Gibson} refers to the apparent motion of objects or features between consecutive frames or views. This module calculates the optical flow between the stereo images and further estimates the feature motion caused by the relative perspective of the binocular camera, enabling the second network to restore photometric alignment semantic information based on the estimated motion information.} Then, a CNN-based fusion network is proposed to fuse the recovered semantics and obtain the restored stereo-images. Moreover, CNN-based channel codecs are deployed at the transmitter and receiver to overcome the channel impairments on the semantic transmission. Our main contributions are as follows,
	\begin{itemize}
		\item[$\bullet$] We propose a DNN-based semantic communication framework for stereo vision-based 3D object detection tasks, in which two sets of semantic codecs are used to extract and process two types of semantic information, and a set of CNN-based channel codecs are used to overcome channel impairments. The proposed framework can improve the accuracy of 3D detection tasks under limited communication bandwidth.
		\item[$\bullet$] We introduce a 2D detection network to identify the objects and propose a set of CNN-based networks, collectively referred to as the key area information network, to extract, compress, and recover the object-related semantic information, providing accurate 2D information for the subsequent 3D object detection task.
		\item[$\bullet$] We design a CNN-based network, termed global information network, to extract semantic information related to photometic alignment from the stereo images. An optical flow-driven recovery network is correspondingly proposed at the receiver to recover the semantic information. The proposed scheme can reduce the loss of the left-right photometric alignment information and improve the accuracy of depth inference.
		\item[$\bullet$] We develop a five-step strategy to train the network with different loss functions for each step. The proposed five-step training method can effectively improve training speed and network performance.
	\end{itemize}
	
	The rest of this paper is organized as follows. Section II describes the system model and framework of our semantic communication system. Section III presents the details of the proposed semantic network. The training strategy is introduced in Section IV. Later, simulation results are discussed in Section V. Section VI concludes this paper.

\section{System Model}
\begin{figure}[htbp]
    	\centering
    	\includegraphics[width=3.45in]{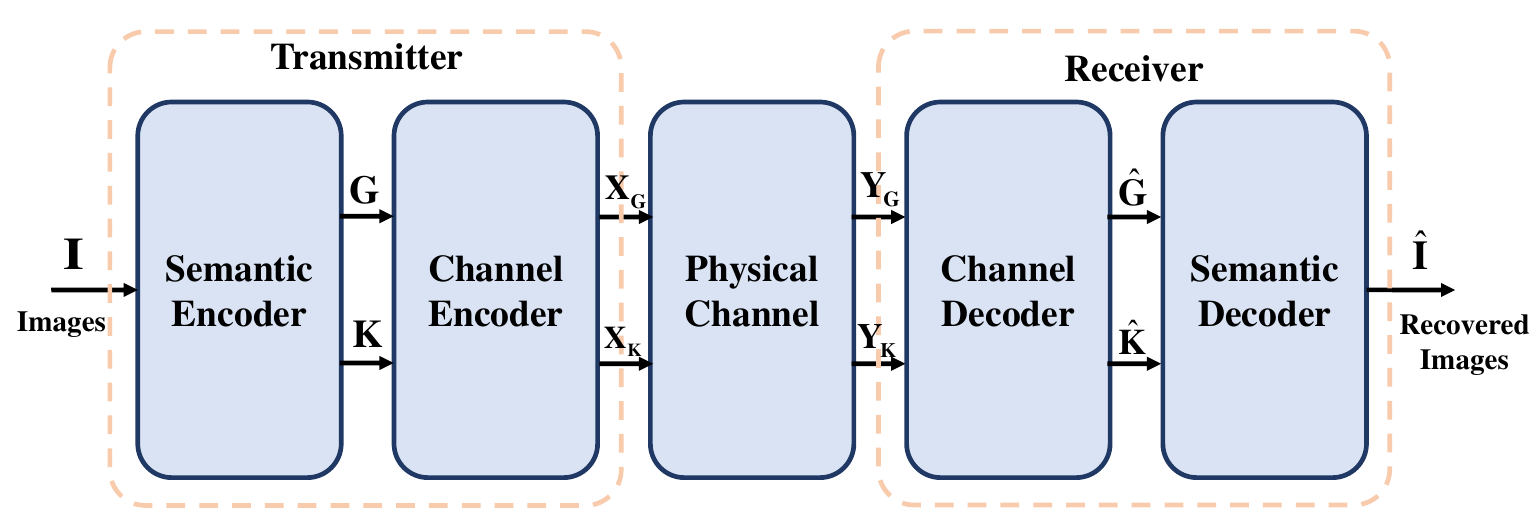}
    	\caption{Model of the proposed semantic communication system.}
    	\label{fig_0}
    \end{figure}

    \textcolor{black}{As shown in Fig.~\ref{fig_0}, the proposed semantic system for stereo-vision 3D object detection tasks adopts the classical architecture similar to \cite{N7}, where the semantic encoder and decoder are responsible for extracting and recovering semantic information, and the channel encoder and decoder are used to mitigate the impacts of wireless channels. The recovered stereo-vision images will be input to another well-trained network for 3D object detection\footnote{\textcolor{black}{In this paper, we restore the original image rather than jointly train the semantic communication network with a specific 3D detection network, aiming to enhance the compatibility of the semantic system with various 3D detection networks.}}. }
    	
    \begin{figure*}[htbp]
    	\centering
    	\includegraphics[width=6.8in]{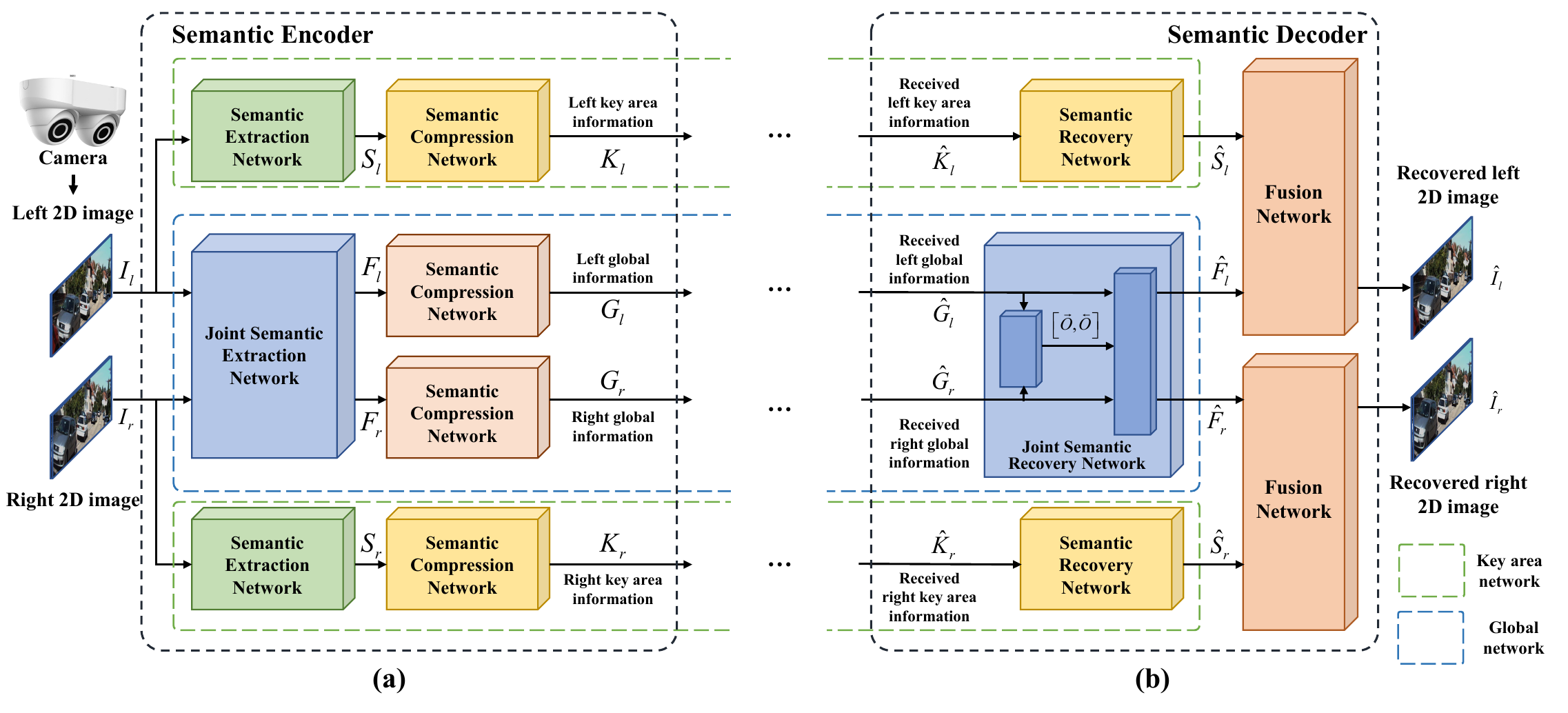}
    	\caption{Illustration of (a) the proposed semantic encoder and (b) the proposed semantic decoder.}
    	\label{fig_1}
    \end{figure*}
    \textcolor{black}{The structure of the semantic codec is shown in Fig.~\ref{fig_1}. Specifically, the proposed semantic codec includes two sets of DNNs to extract, compress, and recover two types of semantic information, i.e., the object-related semantic information for 2D detection and the global semantics containing photometric alignment information for depth inference. Considering the object-related semantic information is exclusively linked to the key area to which objects belong while the second type of semantic information is associated with the images' global information, these two groups of DNNs are referred to as the key area information network and the global information network, respectively. Besides, two fusion networks are used to fuse the recovered semantic information. In the following, we briefly discuss the functions of the networks in both semantic and channel encoders/decoders. The structure of these networks will be presented in detail in Section III.}

\subsection{The Semantic Encoder Network}
    \textcolor{black}{As shown in Fig.~\ref{fig_1}, the semantic encoder network consists of two groups of semantic extraction and compression networks, which correspond to the key area and the global information networks, respectively.}
    
    \textcolor{black}{Specifically, for the key area information network, the semantic extraction networks first use a 2D detection module to estimate the 2D position of vehicles on the left and right images, denoted by $I_l$ and $I_r$, respectively. According to the detected 2D position boxes, the content outside of the boxes is masked while the content within the boxes is extracted as semantic features, which are relevant to 2D object detection. Then, the extracted key area semantic features $\left\{ {{S_l},{S_r}} \right\}$ derived from the left and right images are compressed by the compression networks as the key area information $\left\{ {{K_l},{K_r}} \right\}$, given by
    \begin{equation}
    	\label{eq3_1}
    	\left\{ {\begin{array}{*{20}{c}}
    		{{S_l} = T^l_{\alpha}\left( {{I_l}} \right)}\text{,}\\
    		{{S_r} = T^r_{\alpha}\left( {{I_r}} \right)}\text{,}
    	\end{array}} \right.
    \end{equation}
    and
    \begin{equation}
    	\label{eq3_2}
    	\left\{ {\begin{array}{*{20}{c}}
    		{{K_l} = T^l_{\beta}\left( {{S_l}} \right)}\text{,}\\
    		{{K_r} = T^r_{\beta}\left( {{S_r}} \right)}\text{,}
    	\end{array}} \right.
    \end{equation}
    where $T^l_{\alpha}$ and $T^r_{\alpha}$ denote the semantic extraction networks of the key area information network parameterized by $\alpha$ while $T^l_{\beta}$ and $T^r_{\beta}$ are the compression networks in the key area information network, parameterized by ${\beta}$, to process the semantic features extracted from left and right images, respectively.}
    
    \textcolor{black}{Meanwhile, since the key area information only includes features about the detected objects in a single left or right image, the global information, including photometric alignment information, must also be transmitted to ensure the accuracy of depth inference critical for subsequent 3D detection. Therefore, a joint semantic extraction network is used to jointly extract global semantic features $\left\{ {{F_l},{F_r}} \right\}$ from the stereo-images, where the left and right features are merged to exploit the photometric alignment information between the left and right images. Then, $\left\{ {{F_l},{F_r}} \right\}$ are compressed by the compression networks of the global information network as global information $\left\{ {{G_l},{G_r}} \right\}$, given by
    \begin{equation}
    	\label{eq3_4}
    	\left[F_l, F_r \right] = T_{\gamma}\left(I_l, I_r \right)\text{,}
    \end{equation}
    and
    \begin{equation}
    	\label{eq3_5}
    	\left\{ {\begin{array}{*{20}{c}}
    		G_l = T^l_{\varphi}\left(F_l\right)\text{,}\\
    		G_r = T^r_{\varphi}\left(F_r\right)\text{,}
    	\end{array}} \right.
    \end{equation}
    where $T_{\gamma}$ denotes the joint semantic extraction network parameterized by $\gamma$ while $T^l_{\varphi}$ and $T^r_{\varphi}$ are the compression networks of the global information network, parameterized by ${\varphi}$.}

\subsection{The Channel Encoder and Decoder Networks }
    \textcolor{black}{After the semantic encoder, the channel codec is used to overcome the channel noise. At the transmitter, the channel encoder first encodes the key area and the global information and transmits the encoded information $\left\{ {{X}_K,{X}_G} \right\}$ over the wireless channel, which can be expressed as
    \begin{equation}
    	\label{eq3_00}
    	\left\{ {\begin{array}{*{20}{c}}
    		{X}_K = T_{c_e}\left(K_l, K_r\right)\text{,}\\
    		{X}_G = T_{c_e}\left(G_l, G_r\right)\text{,}
    	\end{array}} \right.	
    \end{equation}
    where $T_{c_e}$ denotes the channel encoder network parameterized by $c_e$. Assuming the transmitted signal undergoes the additive white Gaussian noise (AWGN) channel, the received signal is given by 
    \begin{equation}
    	\label{eq3_01}
    	\left\{ {\begin{array}{*{20}{c}}
    			{Y}_K = {X}_K + {W}_K\text{,}\\
    			{Y}_G = {X}_G + {W}_G\text{,}
    	\end{array}} \right.
    \end{equation}
    where $W_K$ and $W_G$ are the AWGN of the channel. More complicated channel models can be considered as in \cite{N16}.} 
    
    \textcolor{black}{Correspondingly, at the receiver, the channel decoder recovers information $\left\{ {{\hat{K}_r},{\hat{K}_l}} \right\}$ and $\left\{ {{\hat{G}_r},{\hat{G}_l}} \right\}$ from received signals $\left\{ {{Y}_K,{Y}_G} \right\}$, expressed as
    \begin{equation}
    \label{eq3_02}
    	\left\{ {\begin{array}{*{20}{c}}
    		\left\{ {{\hat{K}_l},{\hat{K}_r}} \right\} = T_{c_d}\left(Y_K\right)\text{,}\\
    		\left\{ {{\hat{G}_l},{\hat{G}_r}} \right\} = T_{c_d}\left(Y_G\right)\text{,}
    	\end{array}} \right.
    \end{equation}
    where $T_{c_d}$ denotes the channel decoder network parameterized by $c_d$.}

\subsection{The Semantic Decoder Network}
    \textcolor{black}{Corresponding to the semantic encoder network, the semantic decoder network consists of two sets of semantic recovery networks: the key area information network and the global information network, which are responsible for recovering the key area and global semantic features, respectively. The semantic decoder network also includes two fusion networks to combine these two types of semantic features and obtain the recovered stereo images.}
    
    \textcolor{black}{Specifically, for the key area information network, two recovery networks $T^l_{\theta}(\cdot)$ and $T^r_{\theta}(\cdot)$, parameterized by ${\theta}$, are used to recover key area semantic features $\left\{ {{\hat {S}_l},{\hat{S}_r}} \right\}$ from the received key area information $\left\{ {\hat{K}_l, \hat{K}_r} \right\}$, which can be expressed as
    \begin{equation}
    	\label{eq3_3}
    	\left\{ {\begin{array}{*{20}{c}}
    		\hat{S}_l = T^l_{\theta}\left(\hat{K}_l\right)\text{,}\\
    		\hat{S}_r = T^r_{\theta}\left(\hat{K}_r\right)\text{,}
    	\end{array}} \right.
    \end{equation}
    where $\left\{ {\hat{K}_l, \hat{K}_r} \right\}$ represent the received key area information that is subject to distortion caused by wireless transmission.}
    
    \textcolor{black}{For the global information network, a joint semantic recovery network first uses an optical-flow module to calculate the optical flows $\left\{ {\mathord{\buildrel{\lower3pt\hbox{$\scriptscriptstyle\rightarrow$}} \over {\mathcal{O}}}}, {\mathord{\buildrel{\lower3pt\hbox{$\scriptscriptstyle\leftarrow$}} \over {\mathcal{O}}}} \right\}$ based on the received global information, where $\mathord{\buildrel{\lower3pt\hbox{$\scriptscriptstyle\rightarrow$}} \over {\mathcal{O}}}$ and $\mathord{\buildrel{\lower3pt\hbox{$\scriptscriptstyle\leftarrow$}} \over {\mathcal{O}}}$ represent the optical flows from left-to-right and right-to-left features. Optical flows can effectively describe pixel motions between images or feature maps, assisting in recovering photometric alignment in stereo-vision images. Thus, guided by the optical flows, two CNN modules are then used to recover the global semantic features $\left\{ {{\hat {F}_l},{\hat{F}_r}} \right\}$, given by
    \begin{equation}
    	\label{eq3_6}
    	\left[{\mathord{\buildrel{\lower3pt\hbox{$\scriptscriptstyle\rightarrow$}} \over {\mathcal{O}}}}, {\mathord{\buildrel{\lower3pt\hbox{$\scriptscriptstyle\leftarrow$}} \over {\mathcal{O}}}} \right] = T_{\phi}\left(\hat{G}_l, \hat{G}_r \right)\text{,}
    \end{equation}
    and
    \begin{equation}
    	\label{eq3_7}
    	\left\{ {\begin{array}{*{20}{c}}
    		\hat{F}_l = T^l_{\chi}\left({\mathord{\buildrel{\lower3pt\hbox{$\scriptscriptstyle\leftarrow$}} \over {\mathcal{O}}}}, \hat{G}_l, \hat{G}_r\right)\text{,}\\
    		\hat{F}_r = T^r_{\chi}\left({\mathord{\buildrel{\lower3pt\hbox{$\scriptscriptstyle\rightarrow$}} \over {\mathcal{O}}}}, \hat{G}_r, \hat{G}_l\right)\text{,}
    	\end{array}} \right.
    \end{equation}
    where $\left\{ {\hat{G}_l, \hat{G}_r} \right\}$ represents the received global information, $T_{\phi}$ denotes the optical-flow module parameterized by $\phi$, $T^l_{\chi}$ and $T^r_{\chi}$ are the remaining modules of the recovery networks in the global information network, parameterized by ${\chi}$.}
    
    \textcolor{black}{After recovering the semantic features, two fusion networks in the semantic codec are used to fuse the semantic features and obtain the recovered stereo images $\left\{ {{\hat {I}_l},{\hat{I}_r}} \right\}$, given by
    \begin{equation}
    	\label{eq3_8}
    	\left\{ {\begin{array}{*{20}{c}}
    		\hat{I}_l = T^l_{\nu}\left(\hat{S}_l, \hat{F}_l\right)\text{,}\\
    		\hat{I}_r = T^r_{\nu}\left(\hat{S}_r, \hat{F}_r\right)\text{,}
    	\end{array}} \right.
    \end{equation}
    where $T^l_{\nu}$ and $T^r_{\nu}$ denote the fusion networks, parameterized by ${\nu}$, to recover left and right images, respectively.}
    	
    \textcolor{black}{The final recovered images are fed into a well-trained 3D detection network to estimate the 3D position of the objects. In this paper, the Stereo-RCNN \cite{E14} is used to verify the transmission performance of the proposed communication system, which can also be replaced by other stereo-vision detection networks, e.g., \cite{E5}, \cite{E20}, and \cite{E21}, if needed.}

\section{Transceiver Design for Proposed Semantic Communication System}
	This section describes the detailed structure of the proposed 3D object detection semantic communication system. Based on the system model in section II, the structure is comprised of four parts, the key area information network, the global information network, the fusion network, and the channel codec network. 

	\subsection{Structure of The Key Area Information Network}
	\begin{figure}[htbp]
		\centering
		\includegraphics[width=3.45in]{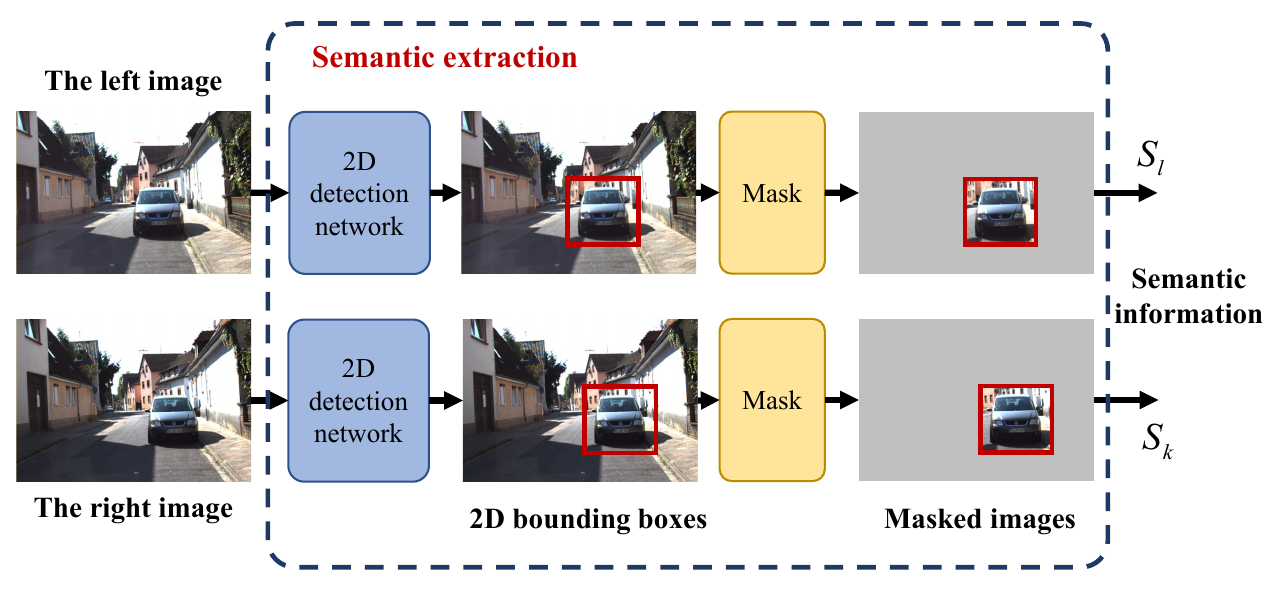}
		\caption{Structure of the semantic extraction network.}
		\label{fig_2}
	\end{figure}
	As shown in Fig.~\ref{fig_2}, a pair of stereo-vision images ${{I_l},{I_r}} \in {\mathbb{R}^{w \times h \times 3}}$ with three RGB channels are first sampled from the dataset $\mathcal{K}$, where $w$ and $h$ represent the width and height of the image. Since the 3D object detection task is mainly related to the image semantics around the object, a 2D detection network is first used to detect the bounding boxes of the objects in the left and right images. Each bounding box has five terms $\left[ u_1, v_1, u_2, v_2, c \right] $, where $\left(u_1, v_1 \right)$ and $\left(u_2, v_2 \right)$ are the coordinates of the upper left and lower right corners of the bounding box, and $c$ is the confidence value, \textcolor{black}{quantifying the predicted probability that the detected region contains an object.} Due to the lightweight requirements of the transmitter, this paper uses a low-complexity network, Yolov5n, as the 2D detection network. Based on the 2D detection results, \textcolor{black}{the information} pertinent to 3D detection, i.e., the image area inside the bounding boxes, is retained while the outside part is masked with zeros. \textcolor{black}{The masked images are regarded as key area semantic features ${{S_l},{S_r}} \in {\mathbb{R}^{w \times h \times 3}}$.} \textcolor{black}{Assuming that the bounding box parameters are transmitted along with the semantic data, the receiver can use these parameters to determine the same mask as the transmitter. Given that the bounding box data is small in amount yet crucial for semantic information reconstruction, similar to control signaling in traditional communication protocols, we assume using a highly reliable transmission scheme to ensure error-free transmission of this data.} By transmitting only the area inside the bounding box along with the very few bounding box parameters, the communication overhead is substantially reduced. This feature extraction process is concisely represented by operation $T^l_{\alpha}(\cdot)$ and $T^r_{\alpha}(\cdot)$ in (\ref{eq3_1}). 
	\begin{figure}[htbp]
		\centering
		\includegraphics[width=3.45in]{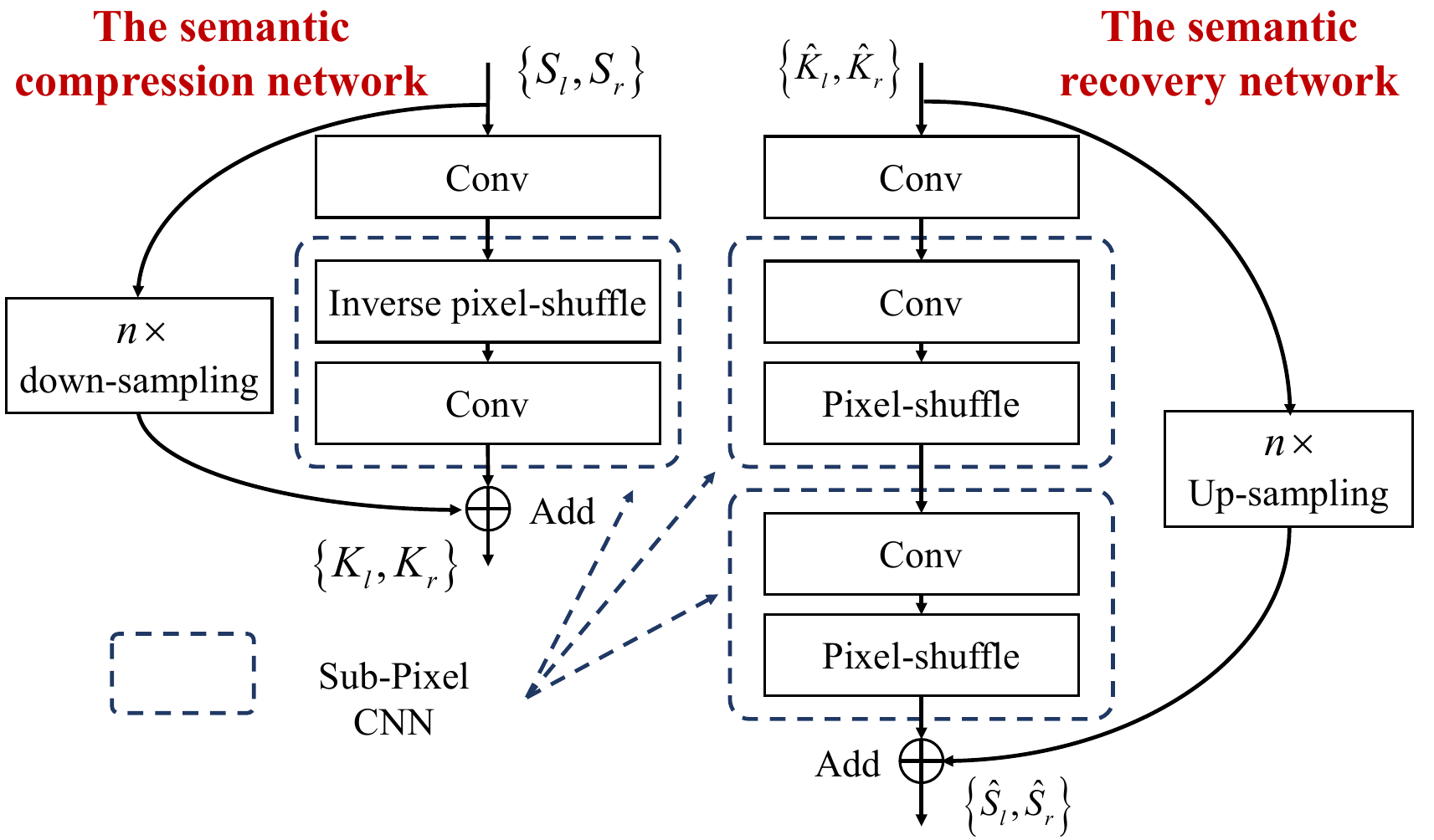}
		\caption{Structure of the semantic compression and recovery module.}
		\label{fig_3}
	\end{figure}
	
	After semantic extraction, two CNN-based networks are used for semantic compression and recovery at the transmitter and receiver, respectively. \textcolor{black}{As shown in Fig.~\ref{fig_3}, the extracted semantic features, $S_l$ and $S_r$, first pass through the sub-pixel CNN to achieve $n \times$ downsampling. The sub-pixel CNN serves as the backbone of the residual network to achieve $n^2$ times compression. The output from the sub-pixel CNN is then combined with the results of $n \times$ bilinear downsampling, which acts as the branch of the residual network, obtaining key area information ${{K_l},{K_r}} \in {\mathbb{R}^{{w/n} \times {h/n} \times 3}}$.} Correspondingly, the recovery network utilizes two sets of sub-pixel CNNs to upsample images progressively. The two sets of upsampling have magnifications of $n_1$ and $n_2$, respectively, where $n_1 \times n_2 = n$. \textcolor{black}{The recovered semantic features} can be expressed as ${{\hat{S}_l},{\hat{S}_r}} \in {\mathbb{R}^{w \times h \times 3}}$. The compression processes correspond to $T^l_{\beta}(\cdot)$ and $T^r_{\beta}(\cdot)$ in (\ref{eq3_2}), and the recovery processes are represented by the operation $T^l_{\theta}(\cdot)$ and $T^r_{\theta}(\cdot)$ in (\ref{eq3_3}).
	\begin{figure}[htbp]
		\centering
		\includegraphics[width=3.45in]{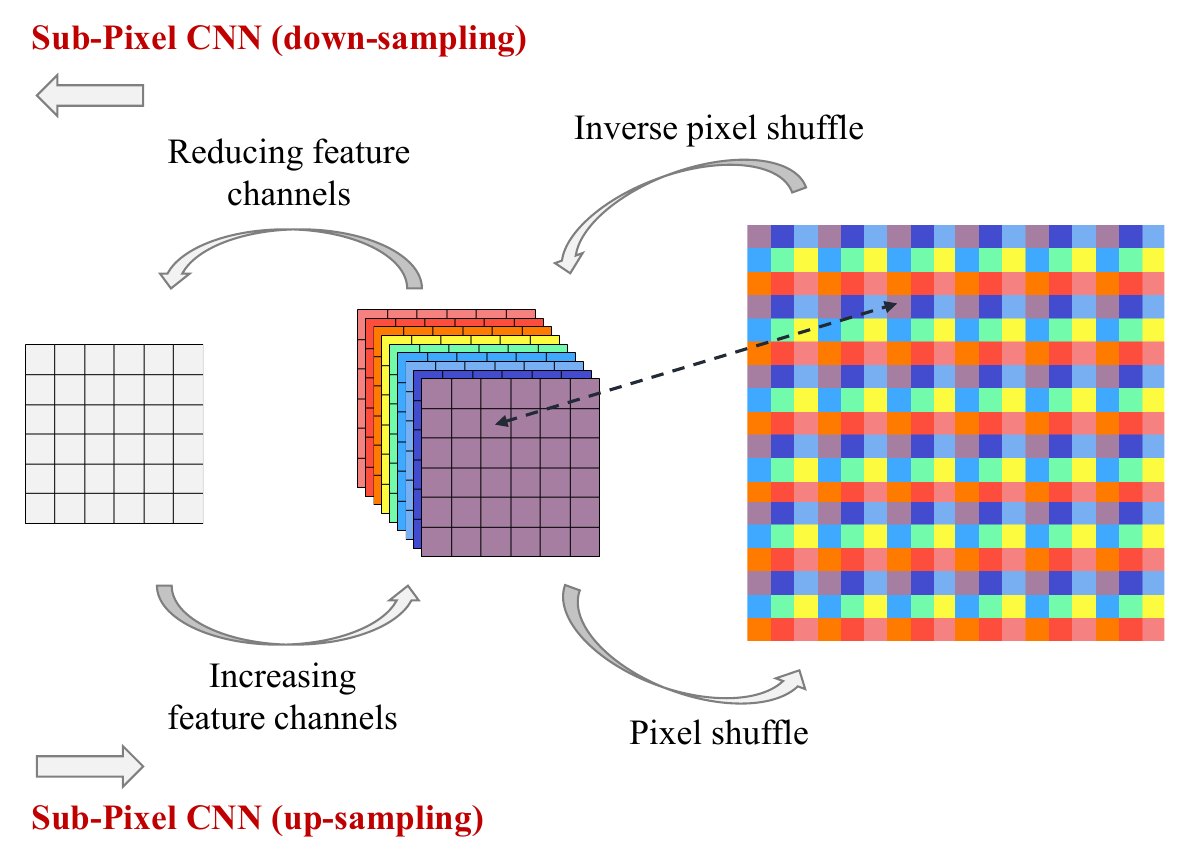}
		\caption{Structure of the sub-pixel CNN.}
		\label{fig_4}
	\end{figure}

	% As shown in Fig.~\ref{fig_4}, the structure of sub-pixel CNNs contains two parts, a CNN module for increasing or reducing the feature channels and a pixel-shuffle module to rearrange the pixels of different channels. Taking $n$ times upsampling as an example (the downsampling can be regarded as the inverse operation of upsampling), for a 2D vector of size $w \times h \times 1$, the sub-pixel CNN first uses multiple convolutional layers to increase its feature channels to $n^2$ and then rearrange the elements in different feature channels to form a new 2D vector of size $nw \times nh \times 1$. Mathematically, the shuffling operation can be described as follows,
	% \begin{equation}
	% 	\label{eq3_9}
	% 	\mathcal{S}{\left( T \right)_{x,y,1}} = {T_{\left\lfloor {x/r} \right\rfloor ,\left\lfloor {y/r} \right\rfloor ,n\bmod \left( {x,n} \right) + n\bmod \left( {y,n} \right) + 1}}\text{,}
	% \end{equation}
	% where $T$ is the output vector of the multiple convolutional layers, $\mathcal{S}$ represents the periodical shuffling operation, and $\left\lfloor \cdot \right\rfloor $ represents the round-down operation. Compared with traditional upsampling methods and the transposed convolution network \cite{E23}, the sub-pixel CNN has better performance and can avoid the uneven overlapping problem.
    As shown in Fig.~\ref{fig_4}, the structure of sub-pixel CNNs contains two parts, a CNN module for increasing or reducing the feature channels and a pixel-shuffle module to rearrange the pixels of different channels. \textcolor{black}{Taking $n$ times upsampling as an example (the downsampling can be regarded as the inverse operation of upsampling), for an input of size $w \times h \times c$, the sub-pixel CNN first uses multiple convolutional layers to increase its feature channels to $n^2c$ and then rearrange the elements in different feature channels to form a new tensor of size $nw \times nh \times c$. Mathematically, the shuffling operation can be described as follows,
    \begin{equation}
        \label{eq3_9}
        \mathcal{S}(\mathbf{T})_{x, y, z} = \mathbf{T}_{\left\lfloor \frac{x}{n} \right\rfloor, \left\lfloor \frac{y}{n} \right\rfloor, (x \bmod n) \cdot n + (y \bmod n) + z \cdot n^2}\text{,}
    \end{equation}}
	where $T$ is the output vector of the multiple convolutional layers, $\mathcal{S}$ represents the periodical shuffling operation, and $\left\lfloor \cdot \right\rfloor $ represents the round-down operation. Compared with traditional upsampling methods and the transposed convolution network \cite{E23}, the sub-pixel CNN has better performance and can avoid the uneven overlapping problem.
	
	\subsection{Structure of The Global Information Network}
	\begin{figure*}[htbp]
		\centering
		\includegraphics[width=6.8in]{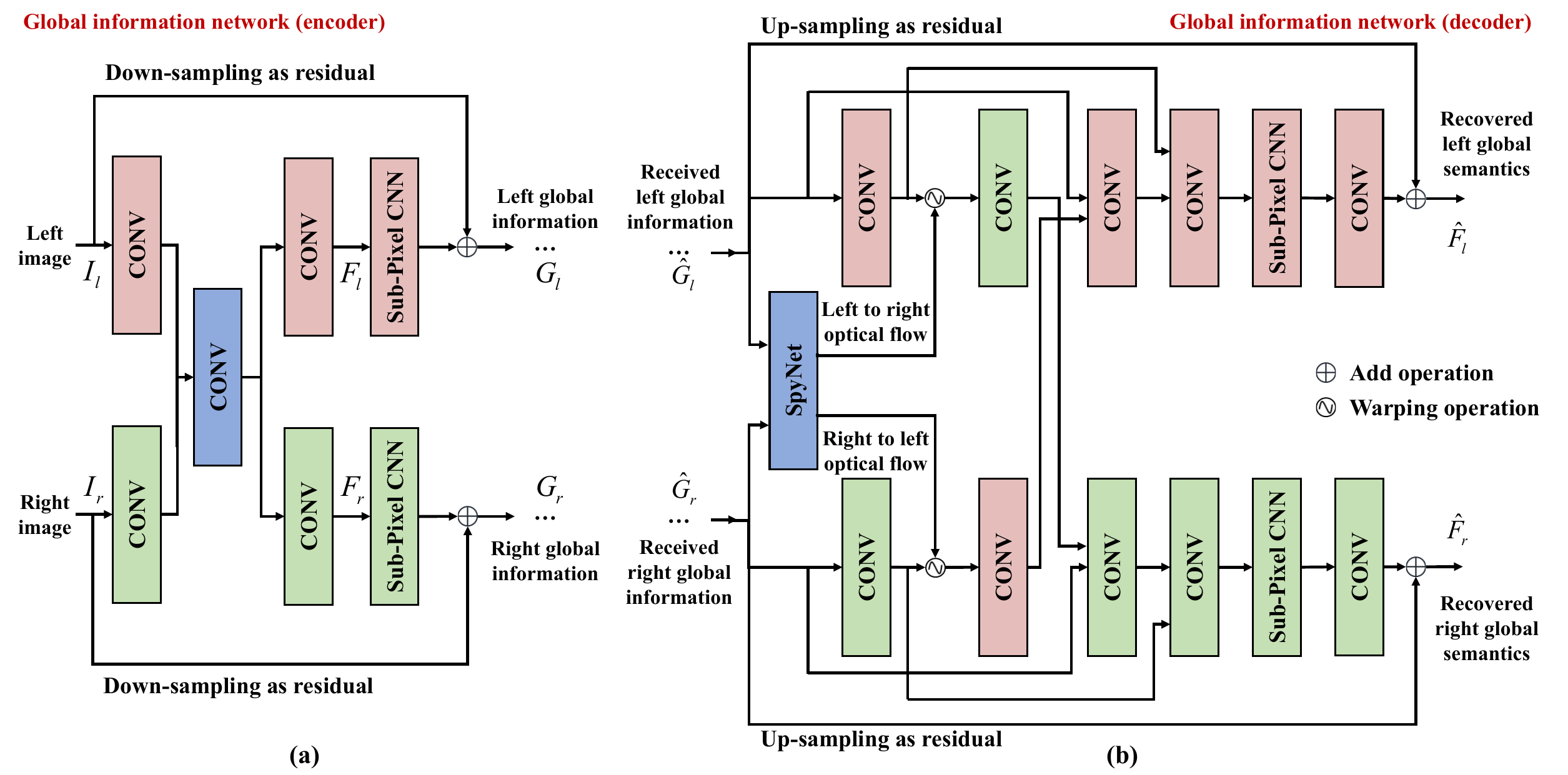}
		\caption{Structure of the global information network (a) in the semantic encoder and (b) in the semantic decoder.}
		\label{fig_5}
	\end{figure*}
	The structure of the global information network is shown in Fig.~\ref{fig_5}. \textcolor{black}{At the transmitter, a convolutional residual network is first utilized to extract features from both the left and right images. These extracted features are combined through direct channel concatenation, which further passes a set of convolutional layers. From the fused features, two sets of convolutional layers with different parameters are then used to separate the left and right semantic features, thereby achieving joint semantic extraction from the left and right images.} The extracted global semantic features are written as $F_l, F_r \in {\mathbb{R}^{w \times h \times f}}$, where $f$ represents the number of feature channels. The semantic features are compressed by the sub-pixel CNN and augmented with the residual information, i.e., the $m$ times downsampling of the original images, obtaining the global information $G_l, G_r \in {\mathbb{R}^{w/m \times h/m \times 3}}$. The feature extraction and compression processes are concisely represented by operation $T_{\gamma}(\cdot)$, $T^l_{\varphi}(\cdot)$ and $T^r_{\varphi}(\cdot)$ in (\ref{eq3_4}) and (\ref{eq3_5}). 
	
	At the receiver, the optical flow method is introduced to utilize the correlation between left and right features for semantic recovery. The optical flow, first proposed by Gibson \cite{Gibson}, describes the projection of a 3D pixel's motion onto a 2D image plane. By calculating the optical flow between two images or feature maps, the relative motion information of the pixels between the images or feature maps can be estimated. Based on the estimated motion information, it can effectively achieve feature alignment between images or feature maps \cite{E24}. Thus, many optical-flow-driven methods are proposed in video super-resolution recovery \cite{E24} and object tracking tasks \cite{E25}. Based on this idea, the receiver adopts the optical flow to align the left and right features and enhance semantic recovery. Besides, a well-trained network, called SpyNet, \cite{E26} is introduced to perform optical flow calculations.
	\begin{figure}[htbp]
		\centering
		\includegraphics[width=3.45in]{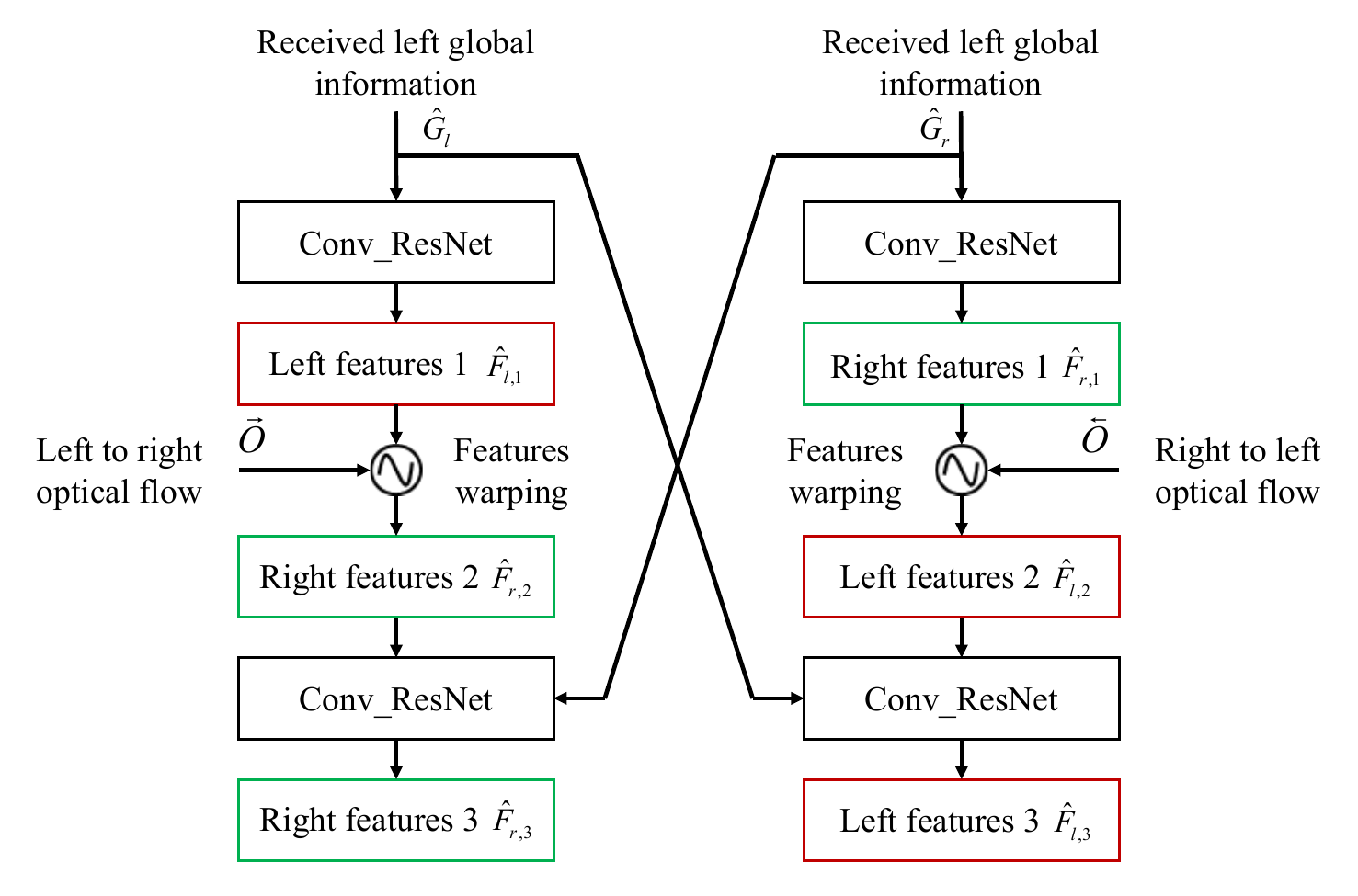}
		\caption{Structure of the optical-flow-driven semantic recovery module.}
		\label{fig_6}
	\end{figure}
	
	The received global information is first input into the SpyNet to estimate the left-to-right and right-to-left optical flow $\left\lbrace \mathord{\buildrel{\lower3pt\hbox{$\scriptscriptstyle\rightarrow$}} \over {\mathcal{O}}}, \mathord{\buildrel{\lower3pt\hbox{$\scriptscriptstyle\leftarrow$}} \over {\mathcal{O}}} \right\} $. \textcolor{black}{Afterward, as shown in Fig.~\ref{fig_6}, the optical-flow-driven recovery module is used to recover the global semantic features, which corresponds to the global information decoder network in Fig.~\ref{fig_5} (excluding SpyNet). Two sets of convolutional residual networks process the received global information $\hat{G}_l, \hat{G}_r$, obtaining feature maps $\hat{F}_{l,1}, \hat{F}_{r,1} \in {\mathbb{R}^{w/m \times h/m \times 3}}$.} The feature maps are then warped by the optical flow to align the features, which can be expressed as follows,
	\begin{equation}
		\label{eq3_10_1}
		\hat{F}_{l, 2} = {\mathcal{F}_{warp}}\left(\hat{F}_{r,1}, \mathord{\buildrel{\lower3pt\hbox{$\scriptscriptstyle\leftarrow$}} \over {\mathcal{O}}} \right)\text{,}
	\end{equation}
	and
	\begin{equation}
		\label{eq3_10_2}
		\hat{F}_{r, 2} = {\mathcal{F}_{warp}}\left(\hat{F}_{l,1}, \mathord{\buildrel{\lower3pt\hbox{$\scriptscriptstyle\rightarrow$}} \over {\mathcal{O}}} \right)\text{,}
	\end{equation}
	where $\mathcal{F}_{warp}$ represents the warping operation and $\hat{F}_{l, 2}, \hat{F}_{r, 2} \in {\mathbb{R}^{w/m \times h/m \times 3}}$ are the warped left and right feature maps. Note that after optical warping, left features are warped to right features and vice versa. After the warping, right features $\hat{F}_{r, 2}$ is concatenated with the received global information $\hat{G}_{r}$ in the feature channels and input into convolutional residual networks, obtaining new feature map $\hat{F}_{r, 3}$. Meanwhile, new left feature map $\hat{F}_{l, 3}$ can be obtained from $\hat{G}_{l}$ and $\hat{F}_{l, 2}$ in the same way. Finally, feature maps $\left\{ \hat{F}_{l, 1}, \hat{F}_{l, 3} \right\}$ and $\left\{ \hat{F}_{r, 1}, \hat{F}_{r, 3} \right\}$ are fused by multiple convolutional layers respectively and recovered by the sub-pixel CNN. The direct upsampling global information is added to the recovered feature maps as the residuals, obtaining \textcolor{black}{recovered global semantic features} $\hat{F}_{l}, \hat{F}_{r} \in {\mathbb{R}^{w \times h \times 3}}$. The optical-flow module corresponds to $T_{{\phi}}(\cdot)$ in (\ref{eq3_6}), and the recovery modules are represented by operation $T^l_{\chi}(\cdot)$ and $T^r_{\chi}(\cdot)$ in (\ref{eq3_7}).

	\subsection{Structure of The Fusion Network}
\begin{figure}[htbp]
		\centering
		\includegraphics[width=3.45in]{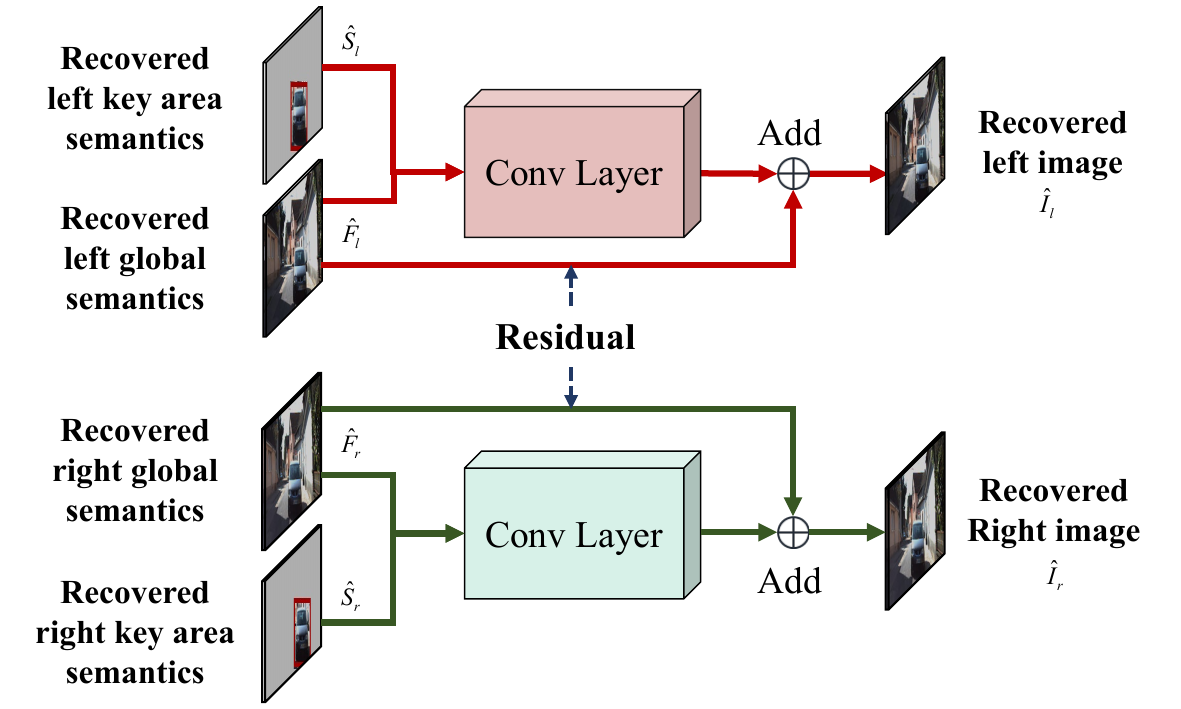}
		\caption{Structure of the fusion network.}
		\label{fig_7}
	\end{figure}

	After the recovery network, the fusion networks combine \textcolor{black}{the key area and the global semantic features} to generate stereo-vision images, as shown in Fig.~\ref{fig_7}. The residual network architecture of the fusion networks fuses \textcolor{black}{semantic features} through multiple convolutional layers and adds features as residuals, obtaining the stereo-vision images. The recovered images are finally input into stereo RCNNs for 3D detection. The fusion networks correspond to $T^l_{\nu}(\cdot)$ and $T^r_{\nu}(\cdot)$ in (\ref{eq3_8}). 

	\subsection{Structure of The Channel Codec Network}
	Multiple convolutional layers form both the channel encoder and the channel decoder, with an exemplary network structure given in Table~\ref{appa2} in Appendix A. The input layer dimension of the channel encoder is consistent with the compressed information dimension while the output layer dimension is adjustable to match the desired channel coding rate. Correspondingly, the channel decoder's input and output layer dimensions are in correspondence to those of the channel encoder. Besides, the dimensions of the intermediate layers of the channel codec are determined by the design requirements.

\section{Training Strategy}
	We propose a five-step training method to train the whole network. The first four steps are used to train the semantic codec network and the last step trains the channel codec network along with the overall end-to-end fine-tuning.

	\textcolor{black}{Step 1 trains the global information network and the fusion network. The training images are sampled from the stereo-vision 3D object detection dataset and pass through the global information network and the fusion network sequentially. Since the key area information network is not trained yet, the recovered key area semantics input into the fusion network are temporarily replaced by all-zero tensors. Besides, a pre-trained SpyNet is used and frozen at the beginning of the training process to train the other part of the global information network. The parameter of the SpyNet is obtained from \cite{E26}. After several epochs of training, the parameter of the SpyNet is unfrozen, and the whole global information network is trained. In Step 1, the Charbonnier loss is used and can be expressed as
	\begin{equation}
		\label{eq5_1}
		\mathcal{L}_{char}({\bf{I}}, {\bf{\hat I}}) = \sqrt {{{\left\| {{\bf{I}} - {\bf{\hat I}}} \right\|}^2} + {\epsilon ^2}}\text{,}
	\end{equation}
	where $\bf{I}$ and $\bf{\hat{I}}$ represent the input stereo-vision images and the output recovered images, respectively, and $\epsilon^2$ is a constant. The pseudocode of the Step 1 is provided in Algorithm \ref{alg1}.}
	\begin{algorithm}
		\caption{Training the global information network}
		\label{alg1}
		\begin{algorithmic}[1]
			\STATE {\bf Input:} The 3D vehicle detection dataset $\mathcal{K}_1$.
			\STATE {\bf Initialize parameters:} $\gamma$, $\varphi$, $\phi$, $\chi$, ${\nu}$. 
			\STATE Load and freeze the pre-trained SpyNet.
			\REPEAT
				\STATE Sample from dataset $\mathcal{K}_1$, obtaining the stereo-vision images $\mathbf{I} \coloneqq \left[I_l, I_r\right]$
				\STATE Extract the semantic features: $\left[F_l, F_r \right] = T_{\gamma}\left(I_l, I_r \right)$
				\STATE Compress the semantic features: \\ \quad \quad $G_l = T^l_{\varphi}\left(F_l\right), G_r = T^r_{\varphi}\left(F_r\right)$
				\STATE Extract the optical flow: $\left[{\mathord{\buildrel{\lower3pt\hbox{$\scriptscriptstyle\rightarrow$}} \over {\mathcal{O}}}}, {\mathord{\buildrel{\lower3pt\hbox{$\scriptscriptstyle\leftarrow$}} \over {\mathcal{O}}}} \right] = T_{\phi}\left(\hat{G}_l, \hat{G}_r \right)$
				\STATE Recover the semantic features: \\ \quad \quad $\hat{F}_l = T^l_{\chi}\left({\mathord{\buildrel{\lower3pt\hbox{$\scriptscriptstyle\leftarrow$}} \over {\mathcal{O}}}}, \hat{G}_l, \hat{G}_r\right), \hat{F}_r = T^r_{\chi}\left({\mathord{\buildrel{\lower3pt\hbox{$\scriptscriptstyle\rightarrow$}} \over {\mathcal{O}}}}, \hat{G}_r, \hat{G}_l\right)$
				\textcolor{black}{\STATE Fuse the semantic features: \\ \quad \quad $\hat{I}_l = T^l_{\nu}\left(\mathbf{0}, \hat{F}_l\right),
				\hat{I}_r = T^r_{\nu}\left(\mathbf{0}, \hat{F}_r\right)$}
				\STATE Optimize $\gamma$, $\varphi$, $\chi$, ${\nu}$ with loss $\mathcal{L}_{char}$ using gradient descent.
			\UNTIL{Reach the preset number of repetitions.}
			\STATE Unfreeze the SpyNet.
			\REPEAT
				\STATE Sample from dataset $\mathcal{K}_1$, obtaining the stereo-vision images $\mathbf{I} \coloneqq \left[I_l, I_r\right]$
				\STATE Extract the semantic features: $\left[F_r, F_l \right] = T_{\gamma}\left(I_r, I_l \right)$
				\STATE Compress the semantic features: \\ \quad \quad $G_l = T^l_{\varphi}\left(F_l\right), G_r = T^r_{\varphi}\left(F_r\right)$
				\STATE Extract the optical flow: $\left[{\mathord{\buildrel{\lower3pt\hbox{$\scriptscriptstyle\rightarrow$}} \over {\mathcal{O}}}}, {\mathord{\buildrel{\lower3pt\hbox{$\scriptscriptstyle\leftarrow$}} \over {\mathcal{O}}}} \right] = T_{\phi}\left(\hat{G}_l, \hat{G}_r \right)$
				\STATE Recover the semantic features: \\ \quad \quad $\hat{F}_l = T^l_{\chi}\left({\mathord{\buildrel{\lower3pt\hbox{$\scriptscriptstyle\leftarrow$}} \over {\mathcal{O}}}}, \hat{G}_l, \hat{G}_r\right), \hat{F}_r = T^r_{\chi}\left({\mathord{\buildrel{\lower3pt\hbox{$\scriptscriptstyle\rightarrow$}} \over {\mathcal{O}}}}, \hat{G}_r, \hat{G}_l\right)$
				\textcolor{black}{\STATE Fuse the semantic features: \\ \quad \quad $\hat{I}_l = T^l_{\nu}\left(\mathbf{0}, \hat{F}_l\right),
				\hat{I}_r = T^r_{\nu}\left(\mathbf{0}, \hat{F}_r\right)$}
				\STATE Optimize $\gamma$, $\varphi$, $\phi$, $\chi$, ${\nu}$ with loss $\mathcal{L}_{char}$ using gradient descent.
				\UNTIL{The loss function converges}
			\STATE {\bf Return:} Network parameters: $\gamma$, $\varphi$, $\phi$, $\chi$, ${\nu}$.
		\end{algorithmic}
	\end{algorithm}
	
	\textcolor{black}{Step 2 successively trains the key area information network, where the 2D detection module, i.e., the YOLOv5n network, is pre-trained and frozen. In this step, the training images are sampled from the same dataset as Step 1 and only pass through the key area information network. Similarly, the Charbonnier loss is used and can be expressed as
	\begin{equation}
		\label{eq5_3_1}
		\mathcal{L}_{char}({\textbf{S}_{mask}}, {\hat{\textbf{S}}_{mask}}) = \sqrt {{{\left\| {{\textbf{S}_{mask}} - {\hat{\textbf{S}}_{mask}}} \right\|}^2} + {\epsilon ^2}}\text{,}
	\end{equation}
	where ${\mathbf{S}_{mask}}$ and ${\mathbf{\hat S}_{mask}}$ represent the masked extracted semantic features at the transmitter and the masked output recovered semantic features at the receiver. The mask is created based on the position boxes of 2D detection to emphasize the neural network's ability to focus on key areas, where the images or features outside the 2D detection boxes are deemed irrelevant. The pseudocode is shown in Algorithm \ref{alg2}.}
	
	In step 3, the fusion network is trained with the frozen pre-trained global and key area information networks. The training dataset is the same as the first part and the loss function is the hybrid masked Charbonnier loss given by
	\begin{equation}
		\label{eq5_3_2}
		\mathcal{L}_{hmc} = \mathcal{L}_{char}\left({\mathbf{I}_{mask}}, {\mathbf{\hat I}_{mask}}\right) + \lambda \mathcal{L}_{char}\left({\bf{I}}, {\bf{\hat I}}\right)\text{,}
	\end{equation}
	where ${\mathbf{I}_{mask}}$ and ${\mathbf{\hat I}_{mask}}$ represent the masked input stereo-vision images and the masked output recovered images. $\lambda$ is a weight used to balance the recovery of the key areas and global areas \textcolor{black}{and is set to $0.5$ in the subsequent simulations.}
	
	\textcolor{black}{In Step 4}, the whole semantic codec except the 2D detection module is trained end-to-end for fine-tuning. \textcolor{black}{During the initial epochs, the training function uses the hybrid masked Charbonnier loss, focusing on recovering key areas. In the final epochs, the training function switches to the Charbonnier loss, emphasizing global recovery.}

	\textcolor{black}{In Step 5}, the channel codec is trained with the semantic codec over the AWGN channel. The training process contains two parts. In the first part, only the channel codec is trained over the AWGN channel with inputs created by the well-trained semantic encoder, where the loss function uses the mean-squared error (MSE) loss \cite{N11, N16}. The second part jointly trains the channel codec and the semantic codec in an end-to-end fashion, where the loss function uses the Charbonnier loss.
    \begin{algorithm}
		\caption{Training the key area information network}
		\label{alg2}
		\begin{algorithmic}[1]
			\STATE {\bf Input:} The 3D vehicle detection dataset $\mathcal{K}_1$.
			\STATE {\bf Initialize parameters:} $\beta$, $\theta$. 
			\STATE {Load and freeze the 2D detection module.}
			\REPEAT
			\STATE Sample from dataset $\mathcal{K}_1$, obtaining the stereo-vision images $\mathbf{I} \coloneqq \left[I_l, I_r\right]$
			\STATE Extract the key area semantic features: \\ \quad \quad $S_l = T^l_{\alpha}\left(I_l\right), S_r = T^r_{\alpha}\left(I_r\right)$
			\STATE Compress the key area semantic features: \\ \quad \quad $K_l = T^l_{\beta}\left(S_l\right), K_r = T^r_{\beta}\left(S_r\right)$
			\STATE Recover the key area semantic features: \\ \quad \quad $\hat{S}_l = T^l_{\theta}\left(\hat{K}_l\right), \hat{S}_r = T^r_{\theta}\left(\hat{K}_r\right)$
			\textcolor{black}{\STATE Optimize $\beta$, $\theta$ with $\mathcal{L}_{char}$ using gradient descent.}
			\UNTIL{Reach the preset number of repetitions.}
			\STATE {\bf Return:} The network parameters: $\beta$, $\theta$.
		\end{algorithmic}
	\end{algorithm}

\section{Simulation Results and Discussions}
%   \subsection{Simulation Settings}	
	In this section, we experimentally evaluate the performance of the proposed semantic communication design for stereo-vision 3D object detection. All methods are evaluated on the KITTI 3D object detection dataset \cite{E27}, where 7481 training images are split into training and validation sets with roughly the same amount. The cars in the images are divided into three regimes with different levels of difficulty: easy, moderate, and hard, according to their 2D box height, occlusion, and truncation levels.  
	
	For the first set of experiments, three average compression ratios are simulated, namely $10 \times$, $30 \times$, and $50 \times$, to verify the network performance at low, medium, and high compression ratios. Perfect communication is assumed between the transmitter and receiver to focus on performance evaluation of semantic encoding/decoding designs. These three compression schemes have different global and key area information compression levels\footnote{\textcolor{black}{Various combinations of compression ratios for key areas and global information are tested during the simulations. The combination with the best detection performance is adopted here.}}. The $10 \times$ scheme compresses the global information $36 \times$ while keeping the key area information uncompressed. The $30 \times$ method compresses the global information $36 \times$ and the key area information $4 \times$. The $50 \times$ compression technique compresses the global information $64 \times$ and the key area information $16 \times$. Correspondingly, the model parameters of the proposed network will also be adjusted. \textcolor{black}{For the second set of experiments, the $30 \times$ source compression scheme is adopted to evaluate the proposed channel codec under AWGN and Rayleigh channels at different SNR levels, ranging from 6 to 18 dB. For the third set of experiments, the proposed semantic communication system is compared with other source-channel coding algorithms over AWGN channels with varying SNRs to evaluate the joint performance of semantic coding and channel coding in the proposed system.}
    
	The following benchmarks are investigated for performance comparison.
	\begin{itemize}
		\item[$\bullet$] \textcolor{black}{ \textbf{\emph{Benchmarks for the first set of experiments:}} We use standard image compression algorithms as benchmarks, including JPEG, JPEG2000, the Super-Resolution Convolutional Neural Network (SRCNN) \cite{E28}, and the Epipolar Cross-Attention Stereo Image Compression (ECSIC) network \cite{ECSIC}. All neural network outputs (except for ECSIC) are quantized to 8-bit and maintain average compression ratios of 10$\times$, 30$\times$, and 50$\times$ by adjusting the output dimensions.}
		\item[$\bullet$] \textbf{\emph{Benchmarks for the second set of experiments:}} Based on the semantic codec network with a $30 \times$ compression ratio, we compare the proposed channel codec network with two traditional transmission schemes, i.e., the 2/3 rate LDPC code with a 64QAM modulation and the 1/2 rate LDPC code with 256QAM modulation. Please note that these two traditional schemes will deliver the same number of source bits and we also adjust the network dimensions of the proposed method such that they all have the same channel uses for fair comparison. 
		\item[$\bullet$] \textcolor{black}{\textbf{\emph{Benchmarks for the third set of experiments:}} We adopt the same settings as the second set of experiments for the proposed method. The comparison approaches employ JPEG, JPEG2000, and ECSIC with $30 \times$ compression as the source coding methods, along with two channel-coding and modulation schemes: 2/3 rate LDPC code with 64QAM modulation and 1/2 rate LDPC code with 256QAM modulation.}
	\end{itemize}
	
	Additionally, Appendixes A and B provide more details on network parameters and training settings for the proposed network.

	\subsection{Performance Metrics}
	The peak signal-to-noise ratio (PSNR) and structural similarity (SSIM) metrics are utilized to evaluate the image recovery performance. The PSNR and SSIM of the images inside and outside the key area boxes are calculated, respectively. The 2D average precision (AP) and the 3D AP metrics are also compared in the experiments to assess the detection accuracy of the proposed approach. These metrics are described in detail subsequently.
	
	The PSNR calculates the MSE between two images to measure their difference. A higher PSNR indicates higher image similarity. Given an original image $I$ and a recovered image $\hat{I}$, the PSNR is defined as
	\begin{equation}
		\label{eq6_1}
		\text{PSNR} = 20 \cdot {\log _{10}}\left( {\frac{{{\mathcal{F}_{\max}}\left( I \right)}}{{{\mathcal{F}_{mse}}\left( {I,\hat I} \right)}}} \right)\text{,}
	\end{equation}
	where ${\mathcal{F}_{\max}}\left(I \right)$ is the maximum pixel value of the image, ${\mathcal{F}_{mse}}\left( {I,\hat I} \right)$ represents the MSE between the original and recovered images. 
	
	The SSIM metric \cite{E29} is more focused on capturing edge and texture similarities, which are essential for mimicking human perception. It calculates the SSIM score by segmenting the image into smaller blocks and evaluating each block's luminance, contrast, and structure scores. The SSIM score can be expressed as 
	\begin{equation}
		\label{eq6_2}
		\text{SSIM} = \frac{1}{N}\sum\nolimits_{n = 1}^N {S_{l,n} \times S_{c,n} \times S_{s,n}}\text{,}
	\end{equation}
	where $N$ is the total number of blocks, and $S_{l,n}$, $S_{c,n}$, $S_{s,n}$ represent the luminance, contrast, and structure scores of the $n$th block, respectively, and can be expressed as
	\begin{align}
		\label{eq6_3}
		\left\{ {\begin{array}{*{20}{l}}
				{S_{l} = \frac{{2{\mu_{x}}{\mu _{y}}}}{{\mu_{x}^2 + \mu_{y}^2}}}\text{,}\\
				{S_{c} = \frac{{2{\sigma_{x}}{\sigma _{y}}}}{{\sigma_{x}^2 + \sigma_{y}^2}}}\text{,}\\
				{S_{s} = \frac{{{\sigma_{xy}}}}{{{\sigma_{x}}{\sigma_{y}}}}}\text{,}\\
		\end{array}} \right.
	\end{align}
	where $x$ and $y$ represent the blocks of the original and recovered images, $\mu_x$ and $\sigma_x$ represent the average and the variance of $x$, and $\sigma_{xy}$ represents the covariance of $x$ and $y$. The SSIM score ranges from 0 to 1, with higher values indicating better image recovery.
	
	% Meanwhile, we use 2D, 3D, and bird's-eye view (BV) AP to measure object detection accuracy. For left and right images, only the 2D detection boxes, whose maximum intersection over union (IoU) with the ground-truth boxes is larger than the preset threshold, are considered true positives \cite{E14}. However, in 3D detection, the estimated 3D box will be regarded as true positives only if the left and right detection boxes meet the IoU threshold conditions simultaneously and the selected left and right ground truth boxes belong to the same object  \cite{E14}. For 2D, 3D, and BV detection, AP ranges from 0 to 100. A higher AP indicates higher detection accuracy. In this simulation, we set the IoU threshold to 0.5 and use the official program approved by KITTI to calculate the detected AP value\footnote{https://www.cvlibs.net/datasets/kitti}.
    \textcolor{black}{Furthermore, 2D, 3D, and bird's-eye view (BV) AP metrics are adopted to evaluate the system's performance in object detection. According to the system model, after the stereo-vision image pairs are recovered, they are fed into a well-trained 3D detection network, i.e., Stereo-RCNN, for 2D and 3D detection. Based on the recovered image pairs, basic 2D detection is first conducted for each image by the Stereo-RCNN, and the 2D detection results from each image are then combined to estimate the 3D bounding boxes of objects, achieving 3D detection. By comparing the output detection results of Stereo-RCNN with the ground-truth boxes, the 2D, 3D, and BV AP metrics are calculated. It is worth noting that, in 2D detection, only the 2D detection boxes, whose maximum intersection over union (IoU) with the ground-truth boxes is larger than the preset threshold, are considered true positives \cite{E14}.} However, in 3D detection, the estimated 3D box will be regarded as true positives only if the left and right detection boxes meet the IoU threshold conditions simultaneously and the selected left and right ground truth boxes belong to the same object \cite{E14}. For 2D, 3D, and BV detection, AP ranges from 0 to 100. A higher AP indicates higher detection accuracy. In this simulation, we set the IoU threshold to 0.5 and use the official program approved by KITTI to calculate the detected AP value\footnote{https://www.cvlibs.net/datasets/kitti}.
	
	\subsection{Simulation Results}
	\begin{table}[htbp]
		\centering
		\caption{Recovery Performance Under Different Compression Ratios}
		\label{tab1}
		\renewcommand\arraystretch{1.4}
		\setlength\tabcolsep{5pt}
		\begin{threeparttable}
			\begin{tabular}{c|c|c|c|c}
				\hline
				{Method} & {PSNR\tnote{g} $\uparrow$}  & {PSNR\tnote{k} $\uparrow$} & {SSIM\tnote{g} $\uparrow$} & {SSIM\tnote{k} $\uparrow$}\\
				\hline
				Proposed Method (10$\times$)  & {27.01dB}        & \textbf{45.12dB} & {0.83}        & \textbf{0.99} \\
				JPEG (10$\times$)             & {28.45dB}        & {29.09dB}        & {0.89}        & 0.93          \\
				JPEG2000 (10$\times$)         & {28.31dB}        & {27.52dB}        & {0.86}        & 0.87          \\
				SRCNN (10$\times$) & \textbf{29.03dB} & {28.32dB}        & \textbf{0.89} & 0.90          \\
				\hline
				Proposed Method (30$\times$)  & \textbf{26.75dB} & \textbf{27.21dB} & \textbf{0.83} & \textbf{0.92} \\
				JPEG (30$\times$)             & 26.17dB          & 25.50dB          & 0.81          & 0.84          \\
				JPEG2000 (30$\times$)         & 26.07dB          & 24.93dB          & 0.78          & 0.78          \\
				SRCNN (30$\times$) & 25.84dB          & 23.56dB          & 0.79          & 0.76          \\
				\hline
				Proposed Method (50$\times$)  & \textbf{25.30dB} & \textbf{23.23dB} & \textbf{0.77} & \textbf{0.81} \\
				JPEG (50$\times$)             & 25.27dB          & 23.05dB          & 0.75          & 0.76          \\
				JPEG2000 (50$\times$)         & 25.05dB          & 21.92dB          & 0.73          & 0.72          \\
				SRCNN (50$\times$) & 24.26dB          & 20.93dB          & 0.74          & 0.69          \\
				\hline
			\end{tabular}
		\end{threeparttable}
		\begin{tablenotes}
			\footnotesize
			\item{g:} the PSNR or SSIM score for the global area.
			\item{k:} the PSNR or SSIM score for the key areas.
		\end{tablenotes}
	\end{table}
	
	For the first set of experiments, Table~\ref{tab1} shows the average PSNR and SSIM scores for different compression ratios across the evaluated methods, where the scores of the key areas (the areas inside the semantic boxes) and the global areas (the whole images) are presented, respectively. All methods use the same validation dataset and transmit similar amounts of data on average. The table shows that the proposed method performs best in key area recovery under the three compression ratios, especially in the SSIM metric. Regarding global performance, the proposed compression scheme performs closely to other algorithms and is slightly better for the medium and high compression ratios.
	\begin{figure*}[htbp]
		\centering
		\includegraphics[width=6.8in]{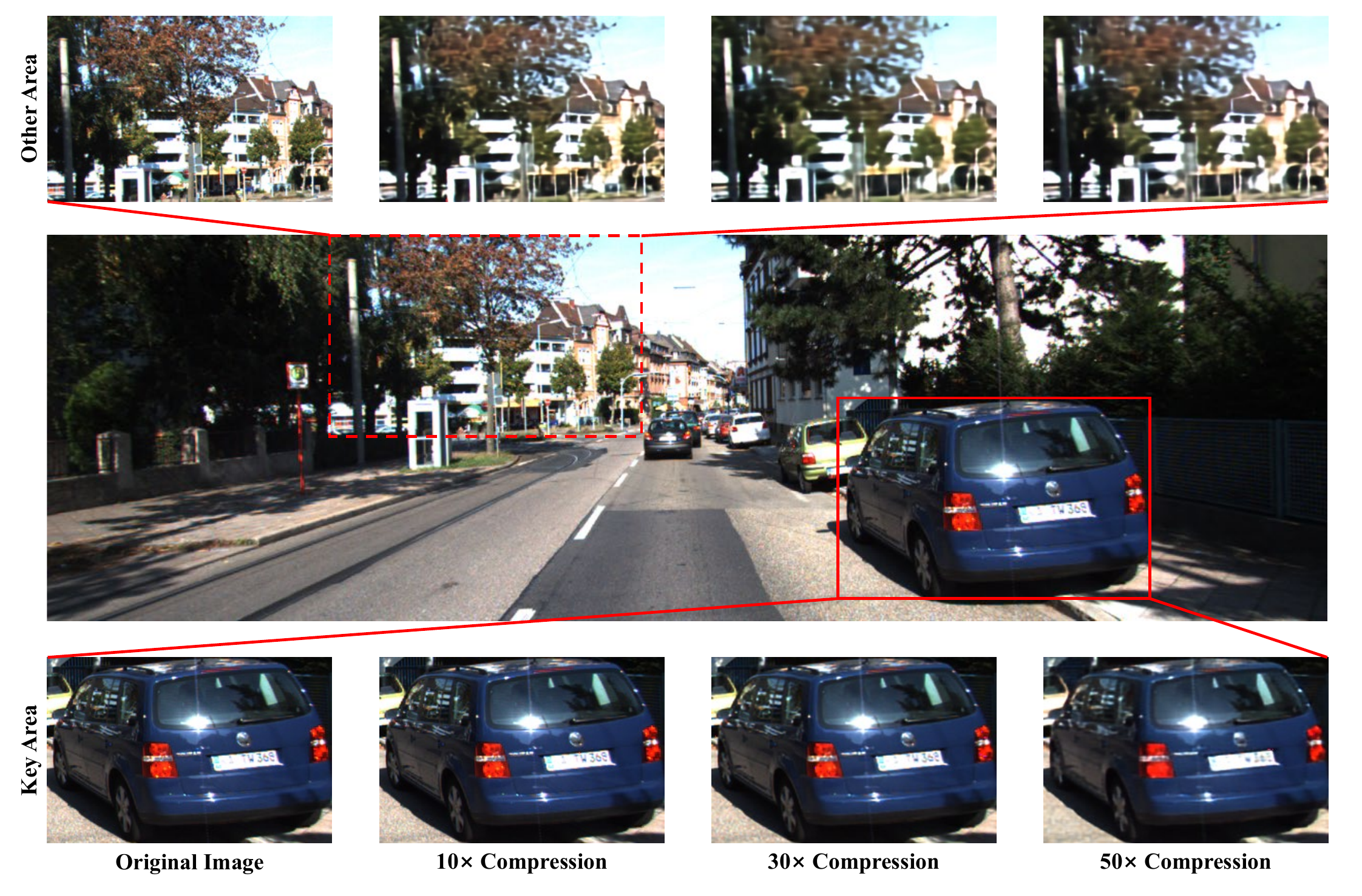}
		\caption{Image recovery results of the proposed method in different areas under different compression ratios.}
		\label{fig_9}
	\end{figure*}
	\begin{figure*}[htbp]
		\centering
		\includegraphics[width=6.8in]{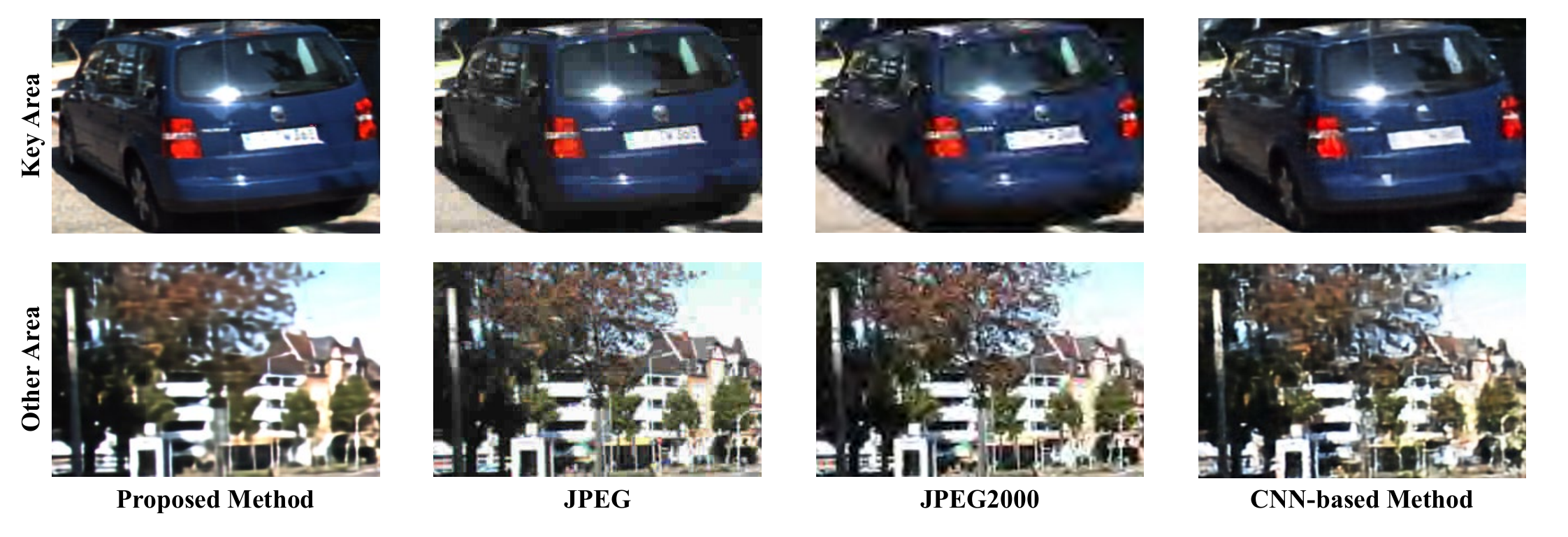}
		\caption{Image recovery results of different methods under $50 \times$ compression ratio.}
		\label{fig_10}
	\end{figure*}
	
	The image recovery results of the proposed method for different compression ratios in key and global areas are shown in Fig.~\ref{fig_9}. \textcolor{black}{The area marked by a solid red box represents the sampled key region, whereas the area enclosed by a dotted red box is considered the sampled other region. By comparing the selected samples from the key and other regions, it can be seen that objects in the key areas retain more distinct outlines and details, although all image qualities decrease as the compression ratio gets higher. Besides, Fig.~\ref{fig_10} presents an example of the recovered images for various methods under the $50 \times$ compression ratio.}

	Continuing with the first set of experiments, the 2D and 3D detection performance in terms of the AP metic with different compression ratios across the evaluated methods is presented in Table~\ref{tab2}. From Table~\ref{tab2}, when compression rates are low, i.e., $10 \times$ compression, the detection performance of all methods is close to that of using the original image directly. This suggests that excessive clarity does not necessarily lead to improved detection accuracy. \textcolor{black}{For medium and high compression rates, the proposed method outperforms all others in 2D and 3D detection metrics, including 2D AP on the left and right images, AP on the bird's eye view, and 3D AP. When the compression rate is $30 \times$, the image recovered by the proposed method has a detection performance similar to the original image for objects in easy and moderate regimes. For objects under the hard regime, the proposed algorithm outperforms other algorithms significantly in 3D detection. Despite achieving a compression ratio of 50 times, the proposed method can still perform similarly to the uncompressed case for objects in easy regimes in terms of 2D AP. Moreover, the proposed algorithm shows slight advantages over the ECSIC method while significantly outperforming other algorithms.}
	\begin{table*}[htbp]
		\centering
		\caption{Detection Performance Under Different Compression Ratios}
		\label{tab2}
		\renewcommand\arraystretch{1.4}
		\setlength\tabcolsep{7pt}
		\begin{tabular}{c|ccc|ccc|ccc|ccc}
			\hline
			& \multicolumn{6}{c|}{\textbf{2D object detection}} & \multicolumn{6}{c}{\textbf{3D object detection}} \\ 
			\cline{2-13} 
			& \multicolumn{3}{c|}{\textbf{AP on left images $\uparrow$}} & \multicolumn{3}{c|}{\textbf{AP on right images $\uparrow$}} & \multicolumn{3}{c|}{\textbf{AP on bird's-eye view $\uparrow$}} & \multicolumn{3}{c}{\textbf{AP for 3D detection $\uparrow$}} \\ 
			\cline{2-13} 
			\multirow{-3}{*}{\textbf{Method}} & \textbf{Easy} & \textbf{Mode} & \textbf{Hard} & \textbf{Easy} & \textbf{Mode} & \textbf{Hard} & \textbf{Easy} & \textbf{Mode} & \textbf{Hard} & \textbf{Easy} & \textbf{Mode} & \textbf{Hard} \\ 
			\hline
			\textbf{Original Image} & \textbf{99.88} & \textbf{90.89} & \textbf{81.80} & \textbf{99.72} & \textbf{90.58} & \textbf{81.38} & \textbf{73.33} & \textbf{47.26} & \textbf{40.94} & \textbf{65.72} & \textbf{45.60} & \textbf{39.39} \\ 
			\hline
			{Proposed Method (10$\times$)}  & {99.85}        & {90.88}          & {81.75}        & {99.66}          & {90.57} & {81.36}  & {73.22} & {47.16} & {39.98} & {{64.94}} & {45.20} & {{39.01}} \\
			{JPEG (10$\times$)}             & {99.79}        & {{90.87}} & \textbf{81.78} & {99.68} & {\textbf{90.58}} & {\textbf{81.38}} & {73.20} & {47.15} & {39.43} & {64.31} & {{45.18}} & {38.74} \\
			{JPEG2000 (10$\times$)}         & {99.78}        & {90.87}          & {81.73}        & {{99.68}} & {90.58} & {81.34}  & {{73.26}} & {{47.19}} & {39.13} & {64.46} & {44.67} & {38.51} \\
			{SRCNN (10$\times$)} & {99.85} & {90.87}          & {81.74}        & {98.80}          & {89.52} & {80.46} & {73.20} & {47.18} & {40.80} & {64.54} & {45.45} & {39.06} \\ 
			{ECSIC (10$\times$)} & \textbf{99.86} & \textbf{90.88}  & {81.78}  & \textbf{99.70} & {90.58} & {81.37} & \textbf{73.30} & \textbf{47.20} & \textbf{40.81} & \textbf{65.44} & \textbf{45.53} & \textbf{39.26} \\ 
			\hline
			{Proposed Method (30$\times$)}  & {\textbf{99.79}} & {\textbf{90.86}} & \textbf{79.71} & {\textbf{99.61}} & {\textbf{90.56}} & {\textbf{79.35}}  & {\textbf{71.06}} & {\textbf{46.72}} & \textbf{39.28} & {\textbf{64.60}} & {\textbf{44.38}} & {\textbf{37.12}} \\
			{JPEG (30$\times$)} & {90.89} & {89.76} & {72.69} & {90.71} & {89.25} & {72.21}  & {65.26} & {45.80} & {35.22} & {61.07} & {39.60} & {33.49} \\
			{JPEG2000 (30$\times$)} & {90.79} & {81.67} & {72.55} & {89.48} & {79.98} & {71.93}  & {63.84} & {40.76} & {34.25} & {55.06} & {37.56} & {31.76} \\
			{SRCNN (30$\times$)} & {90.73} & {81.35} & {63.40} & {88.24} & {77.28} & {60.28} & {65.18} & {42.60} & {35.47} & {55.77} & {34.89} & {28.33} \\  
			{ECSIC (30$\times$)} & {99.74} & {90.78} & {79.58} & {99.54} & {90.37} & {79.09} & {68.82} & {44.28} & {37.39} & {62.68} & {43.49} & {35.41} \\
			\hline
			{Proposed Method (50$\times$)}  & {\textbf{99.67}} & {\textbf{81.88}} & \textbf{72.66} & {{99.13}} & {\textbf{80.85}} & {\textbf{71.71}}  & {\textbf{65.25}} & {\textbf{42.79}} & \textbf{35.87} & {\textbf{60.95}} & {\textbf{40.24}} & {\textbf{33.33}} \\
			{JPEG (50$\times$)} & {71.58} & {62.74} & {53.78} & {70.38} & {61.19} & {52.37}  & {40.49} & {26.60} & {25.66} & {36.92} & {24.50} & {19.59} \\
			{JPEG2000 (50$\times$)} & {81.05} & {71.78} & {62.73} & {77.65} & {67.78} & {59.01}  & {45.37} & {30.14} & {24.56} & {35.44} & {22.46} & {17.42} \\
			{SRCNN (50$\times$)} & {80.14} & {62.40} & {53.43} & {76.35} & {57.99} & {49.37} & {42.26} & {25.11} & {20.09} & {30.99} & {19.53} & {15.51} \\ 
			{ECSIC (50$\times$)} & {99.65} & {81.83} & {72.62} & \textbf{99.14} & {80.07} & {71.60} & {64.33} & {42.56} & {35.63} & {60.18} & {39.89} & {32.64} \\
			\hline
		\end{tabular}
	\end{table*}

\begin{figure*}[htbp]
		\centering
		\includegraphics[width=6.8in]{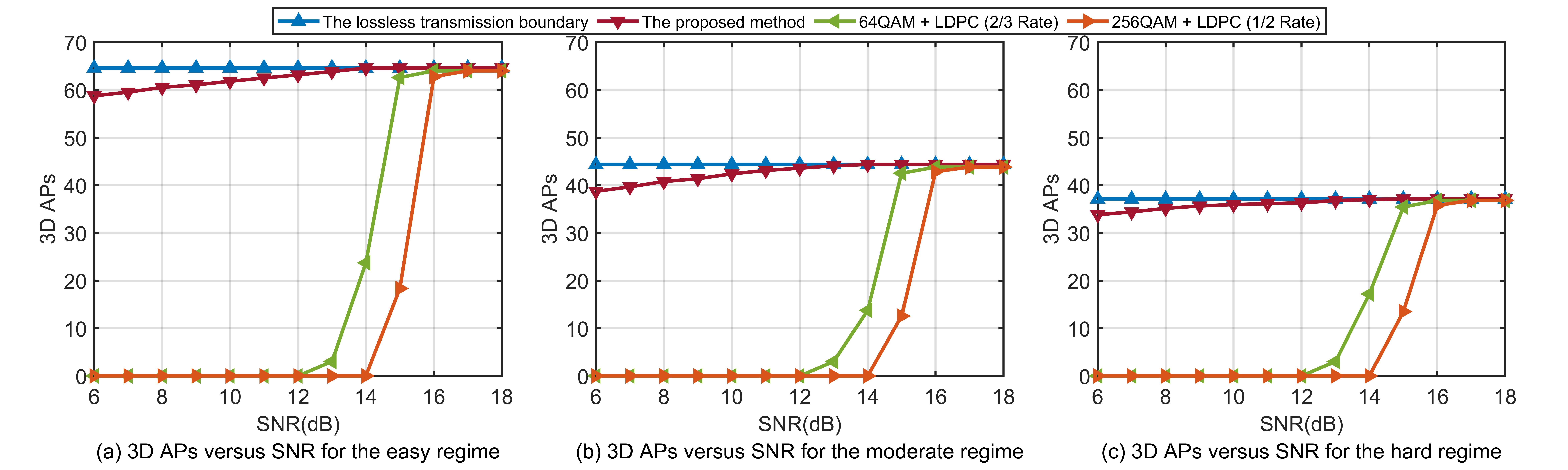}
		\caption{The 3D AP results versus SNR on the AWGN channel for (a) easy, (b) moderate and (c) hard regimes.}
		\label{fig_11}
	\end{figure*}

\begin{figure*}[htbp]
		\centering
		\includegraphics[width=6.8in]{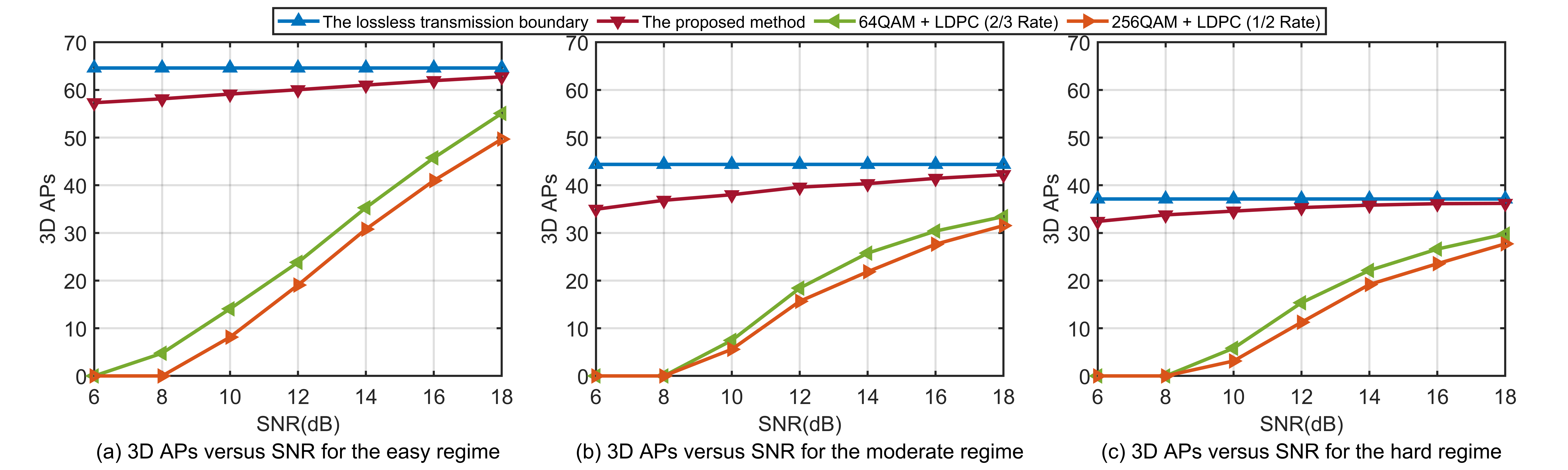}
		\caption{The 3D AP results versus SNR on the Rayleigh fading channel for (a) easy, (b) moderate and (c) hard regimes.}
		\label{fig_a1}
	\end{figure*}

    For the second set of experiments, based on the $30 \times$ compression semantic codec, the proposed channel codec is compared with traditional channel coding and modulation schemes over AWGN and Rayleigh fading channels. \textcolor{black}{In the Rayleigh fading scenario, perfect channel estimation and the zero-forcing equalization algorithm are assumed.} Fig.\ref{fig_11} and Fig.\ref{fig_a1} show the 3D AP results of all methods for different difficulty regimes over AWGN and Rayleigh fading channels, respectively. The traditional schemes use the 2/3 rate LDPC code with 64QAM modulation and the 1/2 rate LDPC code with 256QAM modulation. The output of the semantic codec is quantized to 6-bit. To ensure a fair comparison, the average amount of transferred data remains consistent across all methods. The figures show that the proposed channel codec significantly outperforms the conventional approaches at various SNRs in detection performance over AWGN and Rayleigh fading channels, especially in the low SNR regime.
    
\begin{figure*}[htbp]
		\centering
		\includegraphics[width=6.8in]{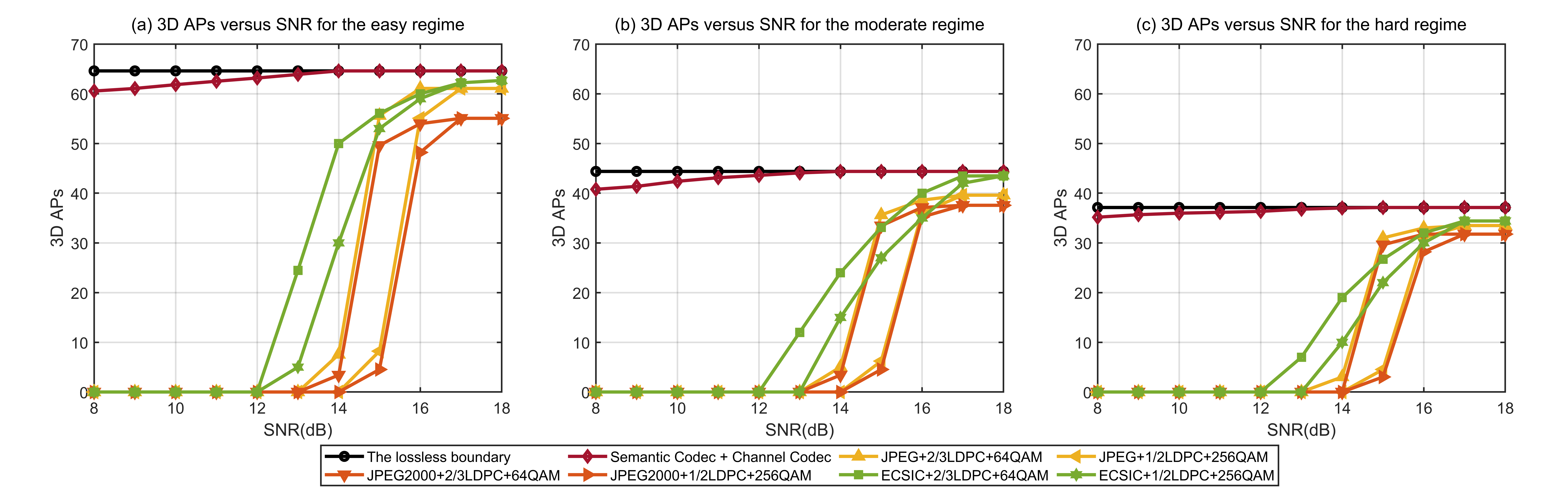}
		\caption{Performance comparison between the proposed semantic communication system and other source-channel coding methods on the AWGN channel for (a) easy, (b) moderate, and (c) hard regimes.}
		\label{fig_a2}
	\end{figure*}

	\textcolor{black}{For the third set of experiments, we evaluate the performance of the proposed semantic communication system compared with other source-channel coding algorithms under AWGN channels at varying SNR levels. The proposed method uses the same settings as in the second set of experiments. The comparison schemes adopt $30 \times$ compressed JPEG, JPEG2000, and ECSIC as source encoding methods, along with two channel coding and modulation methods: the 2/3 rate LDPC code with 64QAM modulation and the 1/2 rate LDPC code with 256QAM modulation. For consistency, all methods transfer the same amount of data. Fig.~\ref{fig_a2} illustrates the performance results of the proposed system and comparison methods across different difficulty regimes. The results demonstrate that the proposed semantic communication system offers superior performance over traditional source-channel coding methods at various SNR levels.}
	
	\subsection{Ablation Study}
	We conduct an ablation study on the proposed network to evaluate the impact of the key area and global information networks on the object detection results, where two networks are separated and work independently. The network separation is simulated by setting the output of the inactive network to zero. For instance, when the global information network is set to work, the output of the key area information network is zero, and vice versa. Besides, the key area and global information are compressed to the same amount to ensure fairness in the experiment. The simulation results of the ablation study are shown in Table~\ref{tab3}.
	% \begin{table}[htpb]
	% 	\centering
	% 	\caption{The 3D Detection Results for The Ablation Study}
	% 	\label{tab3}
	% 	\renewcommand\arraystretch{1.4}
	% 	\setlength\tabcolsep{5pt}
	% 	\begin{threeparttable}
	% 		\begin{tabular}{c|cccccc}
	% 			\hline
	% 			\multirow{3}{*}{{Network}} & \multicolumn{6}{c}{{3D object detection}} \\ 
	% 			\cline{2-7} 
	% 			& \multicolumn{3}{c|}{{AP for bird's-eye view} $\uparrow$}            & \multicolumn{3}{c}{{AP for 3D detection} $\uparrow$} \\ 
	% 			\cline{2-7} 
	% 			& {Easy} & {Mode} & \multicolumn{1}{c|}{{Hard}} & {Easy}    & {Mode}   & {Hard}   \\ 
	% 			\hline
	% 			Network A+B        & \textbf{65.25} &\textbf{42.79} &\multicolumn{1}{c|}{\textbf{35.87}} &\textbf{60.95} &\textbf{40.24} &\textbf{33.33} \\
	% 			Network A\tnote{1} & 25.75          & 22.87         &\multicolumn{1}{c|}{12.28}          & 16.19         & 11.15         & 9.09 \\
	% 			Network B\tnote{2} & 59.39          & 36.51         &\multicolumn{1}{c|}{29.44}          & 50.83         & 29.44         & 22.92 \\ 
	% 			\hline
	% 		\end{tabular}
	% 	\end{threeparttable}
	% 	\begin{tablenotes}
	% 		\footnotesize
	% 		\item{1} Network A denotes the key area information network.
	% 		\item{2} Network B denotes the global information network.
	% 	\end{tablenotes}
	% \end{table}
    \begin{table*}[htbp]
        \centering
        \caption{The Detection Results for The Ablation Study}
        \label{tab3}
		\renewcommand\arraystretch{1.4}
		\setlength\tabcolsep{7pt}
        \begin{threeparttable}
            \begin{tabular}{c|ccc|ccc|ccc|ccc}
                \hline
                & \multicolumn{6}{c|}{\textbf{2D object detection}} & \multicolumn{6}{c}{\textbf{3D object detection}} \\ 
                \cline{2-13} 
                & \multicolumn{3}{c|}{\textbf{AP on left images $\uparrow$}} & \multicolumn{3}{c|}{\textbf{AP on right images $\uparrow$}} & \multicolumn{3}{c|}{\textbf{AP on bird's-eye view $\uparrow$}} & \multicolumn{3}{c}{\textbf{AP for 3D detection $\uparrow$}} \\ 
                \cline{2-13} 
                \multirow{-3}{*}{\textbf{Method}} & \textbf{Easy} & \textbf{Mode} & \textbf{Hard} & \textbf{Easy} & \textbf{Mode} & \textbf{Hard} & \textbf{Easy} & \textbf{Mode} & \textbf{Hard} & \textbf{Easy} & \textbf{Mode} & \textbf{Hard} \\ 
                \hline
                {Network A+B}        & \textbf{99.67} & \textbf{81.88} & \textbf{72.66} & \textbf{99.13} & \textbf{80.85} & \textbf{71.71} & \textbf{65.25} & \textbf{42.79} & \textbf{35.87} & \textbf{60.95} & \textbf{40.24} & \textbf{33.33} \\
                {Network A\tnote{1}} & {83.61} & {67.45} & {51.96} & {81.36} & {65.06} & {50.17} & {38.57} & {25.27} & {24.35} & {27.63} & {20.72} & {16.46} \\
                {Network B\tnote{2}} & {90.68} & {81.06} & {63.27} & {89.25} & {78.40} & {61.21} & {63.77} & {40.92} & {33.73} & {55.07} & {34.29} & {27.84} \\
                \hline
            \end{tabular}
        \end{threeparttable}
        \begin{tablenotes}
            \footnotesize
            \item{1} Network A denotes the key area information network.
            \item{2} Network B denotes the global information network.
        \end{tablenotes}
    \end{table*}
    
	% The table indicates that both key area and global information networks play a significant role in improving the performance of 3D detection. Removing any processing network will result in a substantial loss in overall network performance. The results of the ablation experiment verify the rationality of our network design. The global information provides left-right photometric alignment while the key area information provides object feature details, all required for the 3D object detection task.
    \textcolor{black}{The table indicates that both key area and global information networks play a significant role in improving detection performance. Removing any processing network will result in a substantial loss in overall network performance. The results of the ablation experiment verify the rationality of our network design. The global information network provides left-right photometric alignment to support depth inference, while the key area information network captures object feature details essential for 2D detection, all required for the 3D object detection task.}
	
	\textcolor{black}{Additionally, we perform two ablation studies to assess the effectiveness of the proposed training strategy\footnote{\textcolor{black}{Since training step 5 follows a standard method for training channel codecs, our ablation studies focus on the training strategy for semantic codecs, i.e., training steps 1 to 4.}}. The first study introduces four comparative strategies to evaluate the advantages of the step-by-step training approach. The first comparison strategy combines training steps 2 and 3 of the proposed method. After training the global information network (with its parameters frozen), the key area information network and the fusion network are jointly trained using the hybrid masked Charbonnier loss. In the second strategy, training step 3 is omitted, and joint training is performed after independently training both the global information and key area information networks, utilizing both the hybrid masked Charbonnier loss and the Charbonnier loss. The third strategy jointly trains of all networks following the training of the global information network with two loss functions. Finally, in the fourth strategy, the entire network is trained from the beginning. The 3D detection performance for each strategy is presented in Table~\ref{tab_a2}.}
	\begin{table}[htbp]
		\centering
		\caption{The 3D Detection Results for The Ablation Study 1 of The Training Strategy}
		\label{tab_a2}
		\renewcommand\arraystretch{1.4}
		\setlength\tabcolsep{3.5pt}
		\begin{threeparttable}
			\begin{tabular}{c|cccccc}
				\hline
				\multirow{3}{*}{{Network}} & \multicolumn{6}{c}{{3D object detection}} \\ 
				\cline{2-7} 
				& \multicolumn{3}{c|}{{AP for bird's-eye view} $\uparrow$}            & \multicolumn{3}{c}{{AP for 3D detection} $\uparrow$} \\ 
				\cline{2-7} 
				& {Easy} & {Mode} & \multicolumn{1}{c|}{{Hard}} & {Easy}    & {Mode}   & {Hard}   \\ 
				\hline
				Proposed strategy     & \textbf{71.06} & \textbf{46.72} & \multicolumn{1}{c|}{\textbf{39.28}} & \textbf{64.60} & \textbf{44.38} & \textbf{37.12} \\ 
				Comparison strategy1 & 70.66          & 46.61          & \multicolumn{1}{c|}{39.10}          & 63.80          & 43.96          & 36.23          \\
				Comparison strategy2 & 70.98          & 46.66          & \multicolumn{1}{c|}{39.17}          & 64.47          & 44.16          & 37.01          \\ 
				Comparison strategy3 & 70.59          & 46.60          & \multicolumn{1}{c|}{39.04}          & 63.78          & 43.99          & 36.15          \\ 
				Comparison strategy4 & 67.88          & 44.02          & \multicolumn{1}{c|}{36.89}          & 61.92          & 42.97          & 35.05          \\
				\hline
			\end{tabular}
	    \end{threeparttable}
	\end{table}
	
	\textcolor{black}{The table shows that the performance of the proposed strategy is slightly better than the first to third comparison strategies and significantly better than the fourth comparison strategy, which adopts whole-network training. Additionally, since the proposed method employs a module-by-module training strategy, it can reduce the training time for each epoch. Table~\ref{tab_a3} presents the number of training epochs and the average training time for each step using an Intel Core i5-12400F CPU @ 2.50 GHz and an NVIDIA GeForce GTX 3060. The results demonstrate that the proposed strategy provides the best overall performance with the shortest training time.}
	\begin{table}[htbp]
		\centering
		\caption{The Training Time for The Ablation Study 1 of The Training Strategy}
		\label{tab_a3}
		\renewcommand\arraystretch{1.4}
		\setlength\tabcolsep{3.5pt}
		\begin{threeparttable}
			\begin{tabular}{c|cccc}
				\hline
				Strategy                               & Step 1   & Step 2   & Step 3   & Step 4   \\ \hline
				\multirow{2}{*}{Proposed strategy}     & 12 ep    & 10 ep    & 6 ep     & 10 ep    \\
				& 61 min/ep & 12 min/ep & 32 min/ep & 78 min/ep \\ \hline
				\multirow{2}{*}{Compression strategy1} & 12 ep    & 16ep     & 10 ep    & -        \\
				& 61 min/ep & 39 min/ep & 78 min/ep & -        \\ \hline
				\multirow{2}{*}{Compression strategy2} & 12 ep    & 10 ep    & 16ep     & -        \\
				& 61 min/ep & 12 min/ep & 78 min/ep & -        \\ \hline
				\multirow{2}{*}{Compression strategy3} & 12 ep    & 26 ep    & -        & -        \\
				& 61 min/ep & 78 min/ep & -        & -        \\ \hline
				\multirow{2}{*}{Compression strategy4} & 38 ep     & -        & -        & -        \\
				& 78 min/ep & -        & -        & -        \\ \hline
			\end{tabular}
		\end{threeparttable}
		\begin{tablenotes}
			\footnotesize
			\item{1} ``ep'' denotes the epoch.
		\end{tablenotes}
	\end{table}
	
	\textcolor{black}{We also conduct a second study to assess the necessity of the two loss functions. Two comparison strategies are designed following the same five-step training structure as the proposed approach. The Charbonnier loss is applied to all steps in the first comparison strategy, while the second strategy utilizes the hybrid masked Charbonnier loss throughout. The simulation results in Table~\ref{tab_a4} show that the performance of the proposed method is better than the performance of two comparison strategies, which use either loss function individually.}
	\begin{table}[htbp]
		\centering
		\caption{The 3D Detection Results for The Ablation Study 2 of The Training Strategy}
		\label{tab_a4}
		\renewcommand\arraystretch{1.4}
		\setlength\tabcolsep{3.5pt}
		\begin{threeparttable}
			\begin{tabular}{c|cccccc}
				\hline
				\multirow{3}{*}{{Network}} & \multicolumn{6}{c}{{3D object detection}} \\ 
				\cline{2-7} 
				& \multicolumn{3}{c|}{{AP for bird's-eye view} $\uparrow$}            & \multicolumn{3}{c}{{AP for 3D detection} $\uparrow$} \\ 
				\cline{2-7} 
				& {Easy} & {Mode} & \multicolumn{1}{c|}{{Hard}} & {Easy}    & {Mode}   & {Hard}   \\ 
				\hline
				Proposed strategy    & \textbf{71.06} & \textbf{46.72} & \multicolumn{1}{c|}{\textbf{39.28}} & \textbf{64.60} & \textbf{44.38} & \textbf{37.12} \\ 
				Comparison strategy1 & 68.91          & 44.65          & \multicolumn{1}{c|}{37.81}          & 62.91          & 43.67          & 36.05          \\ 
				Comparison strategy2 & 67.96          & 44.11          & \multicolumn{1}{c|}{37.05}          & 62.03          & 43.11          & 35.37          \\
				\hline
			\end{tabular}
		\end{threeparttable}
	\end{table}
	
	\subsection{Complexity Analysis}
	Table~\ref{tab4} shows the floating-point operations (FLOPs) and parameter counts of neural networks at both the transmitter and receiver. Compared with performing 3D detection at the transmitter, i.e., directly deploying stereo RCNN at the transmitter, or deploying complex feature extraction and optical flow alignment operations at the transmitter, i.e., adding a network close to the scale of “Net B” at the transmitter, the proposed transmitter has lower time and space complexity and is more lightweight, and hardware-friendly. Meanwhile, the previous results show that the proposed semantic communication system can ensure detection accuracy with low communication overhead.
	\begin{table}[htpb]
	 	\centering
	 	\caption{The Complexity Analysis of The Proposed Network}
	 	\label{tab4}
	 	\renewcommand\arraystretch{1.4}
	 	\setlength\tabcolsep{7pt}
	 	\begin{threeparttable}
	 		\begin{tabular}{c|c|c}
	 			\hline
	 			{Network}                          & {Total FLOPs (G)} & {Total Parameters (K)} \\ 
	 			\hline
	 			{Transmitter (Net A)\tnote{1}}        & {0.9}                    & {2.2}                         \\
	 			{Transmitter (Net B)\tnote{2}}        & {176.1}                  & {686.5}                       \\
	 			{Transmitter (Total)}                     & \textbf{177}             & \textbf{688.7}                \\ 
	 			\hline
	 			{Receiver (Net A)\tnote{1}}           & {0.5}                    & {1.6}                         \\
	 			{Receiver (Net B)\tnote{2}}           & {241}                    & {6895.6}                      \\
	 			{Receiver (Net F)\tnote{3}}           & {178.9}                  & {113.1}                       \\
	 			{Receiver (Total)}                        & \textbf{420.4}           & \textbf{7010.3}               \\ 
	 			\hline
	 			{stereo RCNN \tnote{4}}                   & \textbf{448}             & \textbf{108432.4}             \\ 
	 			\hline
	 		\end{tabular}
	 	\end{threeparttable}
	 	\begin{tablenotes}
	 		\footnotesize
	 		\item{1} Net A is the key area information network.
	 		\item{2} Net B is the global information network.
	 		\item{3} Net F is the fusion network.
	 		\item{4} Stereo RCNN is the 3D object detection network.
	 	\end{tablenotes}
 	\end{table}

	\textcolor{black}{The proposed method achieves a trade-off between communication and computation for specific scenarios. Compared with the traditional methods, the semantic-level signal processing in the proposed framework increases the transceiver's computational complexity and computational latency. However, the system also effectively compresses transmitted data and reduces communication latency. This trade-off strategy is particularly suitable for scenarios with limited bandwidth but abundant computing resources at the receiver, such as cloud-car cooperative autonomous driving and cloud-road cooperative monitoring systems.}
 	
\section{Conclusion}	
	This paper proposes a framework for semantic communication for the stereo-vision 3D object detection task that improves 3D detection accuracy under limited communication bandwidth by mainly transmitting the semantic information related to 3D detection. Through the DNN-driven semantic extraction and compression modules, the proposed transmitter can extract two types of semantic information related to 3D detection and compress the semantics at different compression rates to reduce the transmission data. Correspondingly, the receiver also focuses on recovering these two types of semantic information based on the semantic characteristic to reduce the loss of semantic information and maintain the performance of 3D detection tasks. Besides, a CNN-based channel codec is designed to overcome the channel impact of the wireless channel. The simulation results show that the proposed method is more effective in achieving accurate 3D detection but with less data transmission than traditional transmission schemes at various SNRs, especially for the low SNR regime. 

\appendices
\section{The Parameters of The Proposed Network}
This section provides an example of network parameters (Table~\ref{appa1}-\ref{appa3}) with a 30$\times$ compression rate. The compression rate can be adjusted by modifying the parameters of the sub-pixel CNN in the model.
\begin{table}[!ht]
	\centering
	\caption{Model Parameters of The Transmitter Network}
	\label{appa1}
	\renewcommand\arraystretch{1.1}
	\setlength\tabcolsep{10pt}
	\begin{threeparttable}
		\begin{tabular}{ccc}
			\hline
			\multicolumn{1}{c|}{{Module Type}}      & \multicolumn{1}{c|}{{Parameters}}      & {Activation} \\ \hline
			\multicolumn{3}{c}{The Key Area Information Network}                                                                                                 \\ \hline
			\multicolumn{1}{c|}{Conv2d}                    & \multicolumn{1}{c|}{in=3, out=6, kernel=5}    & LeakyReLU           \\
			\multicolumn{1}{c|}{Inv PS}                    & \multicolumn{1}{c|}{in=6, out=24}             & -                   \\
			\multicolumn{1}{c|}{Conv2d}                    & \multicolumn{1}{c|}{in=24, out=3, kernel=3}   & -                   \\ \hline
			\multicolumn{3}{c}{The Global Information Network}                                                                                                       \\ \hline
			\multicolumn{1}{c|}{Conv2d}                    & \multicolumn{1}{c|}{in=3, out=64, kernel=3}   & LeakyReLU           \\
			\multicolumn{1}{c|}{\multirow{2}{*}{ResNet*3}} & \multicolumn{1}{c|}{in=64, out=64, kernel=3}  & ReLU                \\
			\multicolumn{1}{c|}{}                          & \multicolumn{1}{c|}{in=64, out=64, kernel=3}  & -                   \\ \hline
			\multicolumn{1}{c|}{Conv2d}                    & \multicolumn{1}{c|}{in=128, out=64, kernel=1} & LeakyReLU           \\ \hline
			\multicolumn{1}{c|}{Conv2d}                    & \multicolumn{1}{c|}{in=64, out=64, kernel=5}  & LeakyReLU           \\
			\multicolumn{1}{c|}{Conv2d}                    & \multicolumn{1}{c|}{in=64, out=4, kernel=5}   & LeakyReLU           \\
			\multicolumn{1}{c|}{Inv PS}                    & \multicolumn{1}{c|}{in=4, out=144}            & -                   \\
			\multicolumn{1}{c|}{Conv2d}                    & \multicolumn{1}{c|}{in=256, out=3, kernel=3}  & -                   \\ \hline
		\end{tabular}
	\end{threeparttable}
	\begin{tablenotes}
		\footnotesize
		\item{*} ``in" and ``out" represent the input and output channels.
		\item{\#} ``PS" stands for the pixel shuffling.
	\end{tablenotes}
\end{table}

\begin{table}[!ht]
	\centering
	\caption{Model Parameters of The CNN-Based Channel Codec}
	\label{appa2}
	\renewcommand\arraystretch{1.1}
	\setlength\tabcolsep{3pt}
	\begin{threeparttable}
		\begin{tabular}{c|c|c|c}
			\hline
			{Module Type} & {Parameters}     & {Normalization} & {Activation} \\ \hline
			\multicolumn{4}{c}{Transmitter part}                                                                     \\ \hline
			Conv2d                 & in=3, out=64, kernel=3  & -                      & ReLU                \\
			Conv2d $\times 3$      & in=64, out=64, kernel=3 & BatchNorm2d            & ReLU                \\
			Conv2d                 & in=64, out=9, kernel=3  & -                      & -                   \\ \hline
			\multicolumn{4}{c}{Receiver part}                                                                     \\ \hline
			Conv2d                 & in=9, out=64, kernel=3  & -                      & ReLU                \\
			Conv2d $\times 5$      & in=64, out=64, kernel=3 & BatchNorm2d            & ReLU                \\
			Conv2d                 & in=64, out=3, kernel=3  & -                      & -                   \\ \hline
		\end{tabular}
	\end{threeparttable}
	\begin{tablenotes}
		\footnotesize
		\item{*} ``in" represents the input channels, ``out" represents the output channels
	\end{tablenotes}
\end{table}

\begin{table}[!ht]
	\centering
	\caption{Model Parameters of The Receiver Network}
	\label{appa3}
	\renewcommand\arraystretch{1.1}
	\setlength\tabcolsep{10pt}
	\begin{threeparttable}
		\begin{tabular}{ccc}
			\hline
			\multicolumn{1}{c|}{{Module Type}}       & \multicolumn{1}{c|}{{Parameters}}       & {Activation} \\ \hline
			\multicolumn{3}{c}{The Key Area Information Network}                                                \\ \hline
			\multicolumn{1}{c|}{Conv2d}                     & \multicolumn{1}{c|}{in=3, out=24, kernel=3}    & -                   \\
			\multicolumn{1}{c|}{PS}                         & \multicolumn{1}{c|}{in=24, out=6}              & LeakyReLU           \\
			\multicolumn{1}{c|}{Conv2d}                     & \multicolumn{1}{c|}{in=6, out=3, kernel=3}     & LeakyReLU           \\ \hline
			\multicolumn{3}{c}{The Global Information Network}                                                  \\ \hline
			\multicolumn{1}{c|}{Conv2d}                     & \multicolumn{1}{c|}{in=3, out=64, kernel=3}    & LeakyReLU           \\
			\multicolumn{1}{c|}{\multirow{2}{*}{ResNet*30}} & \multicolumn{1}{c|}{in=64, out=64, kernel=3}   & ReLU                \\
			\multicolumn{1}{c|}{}                           & \multicolumn{1}{c|}{in=64, out=64, kernel=3}   & -                   \\
			\multicolumn{1}{c|}{Flow warp}                  & \multicolumn{1}{c|}{in=66, out=64}             & -                   \\
			\multicolumn{1}{c|}{\multirow{2}{*}{ResNet*30}} & \multicolumn{1}{c|}{in=64, out=64, kernel=3}   & ReLU                \\
			\multicolumn{1}{c|}{}                           & \multicolumn{1}{c|}{in=64, out=64, kernel=3}   & -                   \\ \hline
			\multicolumn{1}{c|}{Conv2d}                     & \multicolumn{1}{c|}{in=128, out=64, kernel=1}  & LeakyReLU           \\
			\multicolumn{1}{c|}{Conv2d}                     & \multicolumn{1}{c|}{in=64, out=256, kernel=3}  & -                   \\
			\multicolumn{1}{c|}{PS}                         & \multicolumn{1}{c|}{in=256, out=64}            & LeakyReLU           \\
			\multicolumn{1}{c|}{Conv2d}                     & \multicolumn{1}{c|}{in=64, out=576, kernel=3}  & -                   \\
			\multicolumn{1}{c|}{PS}                         & \multicolumn{1}{c|}{in=576, out=64}            & LeakyReLU           \\
			\multicolumn{1}{c|}{Conv2d}                     & \multicolumn{1}{c|}{in=64, out=64, kernel=3}   & LeakyReLU           \\
			\multicolumn{1}{c|}{Conv2d}                     & \multicolumn{1}{c|}{in=64, out=3, kernel=3}    & -                   \\ \hline
			\multicolumn{3}{c}{The Fusion Network}                                                                  \\ \hline
			\multicolumn{1}{c|}{Conv2d}                     & \multicolumn{1}{c|}{in=6, out=64, kernel=5}    & LeakyReLU           \\
			\multicolumn{1}{c|}{Conv2d}                     & \multicolumn{1}{c|}{in=64, out=64, kernel=5}   & LeakyReLU           \\
			\multicolumn{1}{c|}{Conv2d}                     & \multicolumn{1}{c|}{in=64, out=3, kernel=3}    & -                   \\ \hline
		\end{tabular}
	\end{threeparttable}
	\begin{tablenotes}
		\footnotesize
		\item{*} ``in" and ``out" represent the input and output channels.
		\item{\#} ``PS" stands for pixel shuffling.
	\end{tablenotes}
\end{table}

\section{The Settings of The Proposed Training Strategy}
The detailed parameter settings of the proposed training strategy are shown in Table~\ref{appb1}, where "Network A" represents the key area information network and "Network B" represents the global information network. The optimization algorithm of all steps adopts the Adam algorithm. 
\begin{table}[!ht]
	\centering
	\caption{Parameters of The Training Strategy}
	\label{appb1}
	\renewcommand\arraystretch{1.1}
	\setlength\tabcolsep{3pt}
	\begin{threeparttable}
		\begin{tabular}{c|cc|c}
			\hline
			Step No. & \multicolumn{2}{c|}{Learning Rate}                                                                                                                                                                                                                                                                                                                                                    & Epochs \\ \hline
			Step 1   & \multicolumn{1}{c|}{\begin{tabular}[c]{@{}c@{}}Network B in Transmitter\\ Network B in Receiver\\ SpyNet\\ Fusion Network\end{tabular}}                                                                            & \begin{tabular}[c]{@{}c@{}}$2 \times 10^{-4}$\\ $2 \times 10^{-4}$\\ $2.5 \times 10^{-5}$\\ $10^{-4}$\end{tabular}                                               & 2+10\tnote{*}   \\ \hline
			Step 2   & \multicolumn{1}{c|}{\begin{tabular}[c]{@{}c@{}}Network A in Transmitter\\ Network A in Receiver\end{tabular}}                                                                                                      & \begin{tabular}[c]{@{}c@{}}$10^{-4}$\\ $2 \times 10^{-4}$\end{tabular}                                                                                                    & 10     \\ \hline
			Step 3   & \multicolumn{1}{c|}{Fusion Network}                                                                                                                                                                                & $10^{-4}$                                                                                                                                                        & 6      \\ \hline
			Step 4   & \multicolumn{1}{c|}{\begin{tabular}[c]{@{}c@{}}Network B in Transmitter\\ Network B in Receiver\\ Network A in Transmitter\\ Network A in Receiver\\ SpyNet\\ Fusion Network\end{tabular}}                         & \begin{tabular}[c]{@{}c@{}}$10^{-4}$\\ $10^{-4}$\\ $2 \times 10^{-5}$\\ $2 \times 10^{-5}$\\ $1.25 \times 10^{-5}$\\ $2 \times 10^{-5}$\end{tabular}             & 10     \\ \hline
			Step 5   & \multicolumn{1}{c|}{\begin{tabular}[c]{@{}c@{}}Channel Codec Network\\ Network B in Transmitter\\ Network B in Receiver\\ Network A in Transmitter\\ Network A in Receiver\\ SpyNet\\ Fusion Network\end{tabular}} & \begin{tabular}[c]{@{}c@{}}$10^{-4}$\\ $10^{-4}$\\ $10^{-4}$\\ $2 \times 10^{-5}$\\ $2 \times 10^{-5}$\\ $1.25 \times 10^{-5}$\\ $2 \times 10^{-5}$\end{tabular} & 25+20\tnote{\#}  \\ \hline
		\end{tabular}
	\end{threeparttable}
	\begin{tablenotes}
		\footnotesize
		\item{*} 2 epochs with the frozen SpyNet, 10 epochs with the unfrozen SpyNet.
		\item{\#} 25 epochs for only the channel codec, 20 epochs for the whole network.
	\end{tablenotes}
\end{table}

\bibliographystyle{IEEEtran}
\bibliography{IEEEabrv,main}

%\newpage
%\begin{IEEEbiography}[{\includegraphics[width=1in,height=1.25in,clip,keepaspectratio]{figure/fig1}}]{Michael Shell}
%Use $\backslash${\tt{begin\{IEEEbiography\}}} and then for the 1st argument use $\backslash${\tt{includegraphics}} to declare and link the author photo.
%Use the author name as the 3rd argument followed by the biography text.
%\end{IEEEbiography}
%
%\begin{IEEEbiography}[{\includegraphics[width=1in,height=1.25in,clip,keepaspectratio]{figure/fig1}}]{Michael Shell}
%	Use $\backslash${\tt{begin\{IEEEbiography\}}} and then for the 1st argument use $\backslash${\tt{includegraphics}} to declare and link the author photo.
%	Use the author name as the 3rd argument followed by the biography text.
%\end{IEEEbiography}
%%\begin{IEEEbiographynophoto}{John Doe}
%%Use $\backslash${\tt{begin\{IEEEbiographynophoto\}}} and the author name as the argument followed by the biography text.
%%\end{IEEEbiographynophoto}

\vfill

\end{document}